\documentclass[twocolumn, pra, amssymb, superscriptaddress, aps, showpacs,preprintnumbers,
amsmath,showkeys,floatfix]{revtex4}

\usepackage{epstopdf}
\usepackage{graphics}
\usepackage{graphicx}
\usepackage{dcolumn}
\usepackage{bm}
\usepackage{longtable}
\usepackage{epsfig}
\usepackage{times}
\usepackage{url}
\usepackage{color}

\begin{document}
\title{Transition between Amplified Spontaneous Emission and Superfluorescence in a longitudinally pumped medium by an X-ray free electron laser pulse}
 
\author{Yu-Hung   \surname{Kuan}}
%
\author{Wen-Te \surname{Liao}}
\email{wente.liao@g.ncu.edu.tw}
\affiliation{Department of Physics, National Central University, Taoyuan City 32001, Taiwan}
\date{\today}
\begin{abstract}
The transition from the amplification of spontaneous emission to superfluorescence in a three-level and swept-gain medium excited by an X-ray free electron laser pulse is theoretically investigated. 
Given the specific time scale of X-ray free electron laser pulse, we investigate the swept pumping process in detail and our results  show that the temporal structure of an X-ray free electron laser pulse plays a more critical role than its peak intensity does for producing  population inversion. 
The typical watershed of two radiant regions depends on the optical depth of the gain medium for a given coherence time, namely, particle number density and the medium length are equally important.
However, we find that medium length plays more important role than particle density does for  making the forward-backward  asymmetry. The transient gain length and the total medium length are identified as two important factors to observe length induced backward transition.
The present results suggest an application of  parametric controls over a single-pass-amplified light source.
\end{abstract}

\maketitle
\section{Introduction}
Absorption, stimulated emission and spontaneous emission of photons are three fundamental processes of light-matter interaction. However, a collection of excited atoms emits light differently from what a single atom does. 
One type of  such collective light emission was first theoretically studied in Dicke's pioneering and seminal work of superfluorescence (SF) \cite{Dicke1954}. 
While in free space a wavepacket of photons isotropically emitted by a single atom behaves like it is exponentially decaying with a duration of an excited state lifetime, SF from multiple atoms with particle density $n$ for a given sample length $L$ acts as a directional light burst characterized by its peak intensity, duration and time delay, which are proportional to  $n^2$, $n^{-1}$ and $n^{-1}$, respectively  \cite{Polder1979, Gross1982}.
SF was first experimentally realized in room-temperature HF gas \cite{Skribanowitz1973}, and subsequently observed in, e.g., Sodium \cite{Gross1976}, Cesium \cite{Gibbs1977},  Rubidium atoms \cite{Paradis2008}, solid-state KCl$:$O$_2^-$ \cite{Florian1984, Boyd1987} and  in plasma \cite{Dreher2004, Noe2012}. 
More recently, the development of a free electron laser (FEL) \cite{Deacon1977} has lead to further advancements, such as the observation of EUV-FEL-induced SF in Helium gas \cite{Nagasono2011}, XUV-FEL-induced SF in Xenon gas \cite{Mercadier2019} and FEL superfluorescence \cite{Watanabe2007}.
Apart from typical three-level-$\Lambda$ type atomic systems \cite{Gibbs1977, Nagasono2011, Cui2012, Cui2013, Cui2017}, a V-type one \cite{Keitel1992, Kozlov1999} has also been theoretically investigated.

Another kind of collective  emission is the so-called amplification of spontaneous emission (ASE) \cite{Gross1982, Pert1994, Rohringer2012, Weninger2014, Yoneda2015}, namely, radiation due to spontaneous emission from a single emitter is amplified when it propagates through an excited  sample. While ASE is also directional due to the geometry of the inverted medium, the time structure of ASE is quite distinct from that of SF, e.g., the peak intensity does not behave as $n^2$ \cite{Schuurmans1979, Boyd1987}. 
Utilizing the ASE mechanism, an atomic inner shell laser pumped by an X-ray Free Electron Laser (XFEL) has recently been achieved \cite{Rohringer2012, Weninger2014, Yoneda2015}.
While this achievement \cite{Rohringer2012, Yoneda2015} has become milestone within the field of X-Ray quantum optics \cite{Shvydko1996, Roehlsberger2010, Liao2011, Roehlsberger2012, Liao2012a, Adams2013, Liao2013, Vagizov2014, Liao2014a, Heeg2015a, Liao2015, Heeg2017, Wang2018}, 
XFEL-induced SF  has never been studied.
Moreover, SF and ASE are independently investigated in most works with only  few examples touching on both  \cite{Bonifacio1971a, Bonifacio1971b, Okada1978, Brechignac1981, Boyd1987, Rai1992}. 
Studying XFEL-pumped SF, as well as the transition from ASE to SF using XFEL, is therefore  interesting and timely. 

Most  theoretical studies of ASE or SF rely on three assumptions:
(i) a completely inverted medium as the initial condition \cite{Haake1979, Haake1979b, Schuurmans1979, Gross1982},
(ii) a swept-gain amplifier excited by an oversimplified $\delta$-function pulse  \cite{Arecchi1970, Bonifacio1975, Hopf1975}, and
(iii) only forward emission of light \cite{Arecchi1970, Bonifacio1975, Haake1979, Haake1979b, Schuurmans1979, Rai1992, Weninger2014}.
In view  that XFEL pulse duration may be greater than the excited state lifetime  \cite{Rohringer2012, Yoneda2015}, the assumption (ii) \cite{Arecchi1970, Bonifacio1975, Hopf1975} is unrealistic. One therefore has to carefully deal with the pumping process.
We  use 
(a) a swept-gain amplifier excited by a short XFEL pumping pulse with a duration of a few tens femtosecond \cite{EuropeanXFEL}, and
(b) a set of equations describing both the forward and backward emission.
Our detailed study of this issue identifies the mechanism breaking the  forward and backward symmetry and demonstrates that the time structure of the pumping pulse is highly critical for producing population inversion. 
Typically, the ASE-SF transition depends only on  $\tau_2\gg \sqrt{\tau_R \tau_D}$ or $\tau_R\ll\tau_2\ll \sqrt{\tau_R \tau_D}$, where $\tau_2$ is the coherence time of the  transition, $\tau_R$ the characteristic duration  of SF and $\tau_D$ the SF delay time \cite{Friedberg1976, Schuurmans1979, Boyd1987}. The definition of $\tau_R$ and $\tau_D$ shows that the length of a medium and  particle density are equally important for ASE-SF transition.
However, our study on the pumping process shows that the choice of medium length plays a more important role than particle density does for observing backward emission, namely,  medium length induced ASE-SF transition can happen to the backward emission.
Apart from the typical averaged temporal behavior of emitted light pulses, we show that both the averaged spectrum and the histogram of emitted photon number  manifest the ASE-SF transition. Our results therefore give useful hints for quantifying the XFEL-pumped light source and demonstrate in what parameter region the transition may occur.
The present results suggest an application of modifying the properties of a single-pass light source \cite{Rohringer2012, Weninger2014, Yoneda2015, Gunst2014, Brinke2013, Zhang2013} via the transition between ASE and SF  induced by the change of optical depth of the gain medium or by the variety of a pumping  pulse.

This paper is organized as follows. 
In Sec. \ref{sec:model}, we describe our system and theoretical model using the Maxwell-Bloch equation. 
In Sec. \ref{sec:as}, we present our analysis of the production of population inversion.
In Sec. \ref{sec:nr}, we numerically solve these coupled equations and discuss the transition between ASE and SF. 
In Sec. \ref{sec:fba}, the length effects for forward-backward asymmetry are discussed.
In Sec. \ref{sec:Jptransition}, we demonstrate the transition between ASE and SF induced by the variety of pumping laser parameters. 
A summary is present in Sec. \ref{sec:summary}.

\begin{figure}[b]
\vspace{-0.4cm}
  \includegraphics[width=0.48\textwidth]{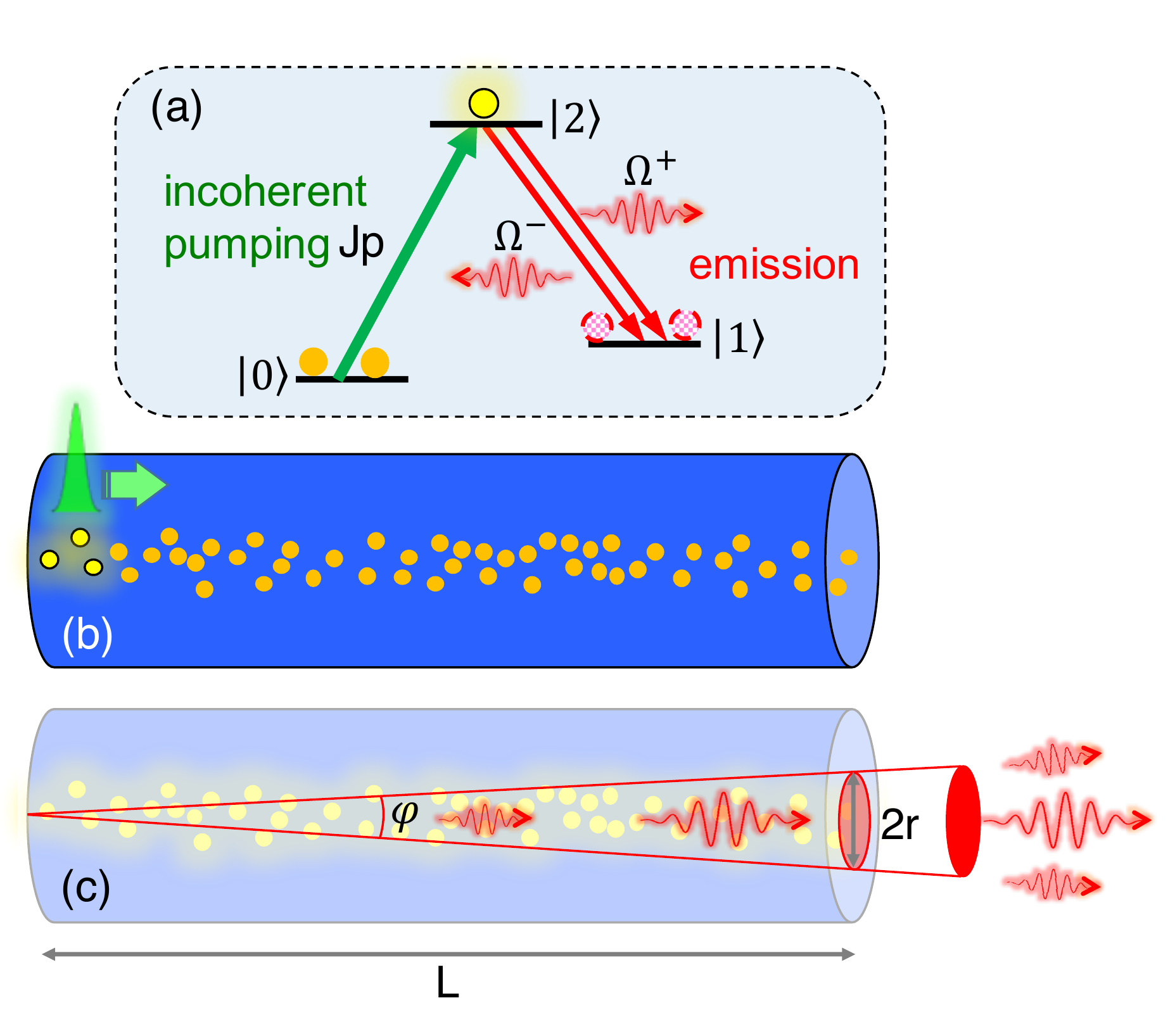}
  \caption{\label{fig1}
(Color online) (a) three-level $\Lambda$-type scheme. Transition $\vert 0\rangle \rightarrow \vert 2\rangle$ is incoherently pumped by a short pumping laser pulse $J_p$ (green upward arrow).  Atoms promoted to state $\vert 2\rangle$ subsequently decay and become state $\vert 1\rangle$. Atoms that undergo the transition $\vert 2\rangle \rightarrow \vert 1\rangle$ emit photons (red downward arrows) in the forward and backward direction with equal probability.
(b) atoms in state $\vert 0\rangle$ (orange dots) are initially prepared in a gas cell. When a pumping pulse (green filled Gaussian pulse) is travelling through, atoms are promoted to state $\vert 2\rangle$ (yellow filled circles) in a region defined by the spot size and path of the pumping pulse $J_p$.
(c) $\varphi$ in Eq.~(\ref{eq16}) represents the solid angle within which the emitted photons due to transition $\vert 2\rangle \rightarrow \vert 1\rangle$ get amplified in the forward direction. $\varphi$ is determined by the length $L$ and the transverse radius $r$ of the medium.
  }
\end{figure}

\section{Model}\label{sec:model}
\begin{table}[t]
\centering
\caption{\label{table1}
The notation used throughout the text. The indices $i, j \in \left\lbrace 0, 1, 2\right\rbrace $ denote the three states showed in Fig.~\ref{fig1}(a). 
}
\begin{tabular}{cl}
\hline
\hline
Notation & \multicolumn{1}{c}{Explanation} \\
\hline
$c$ &  Speed of light in vacuum. \\
$\varepsilon_0$ &   Vacuum permittivity. \\
$\Omega^{+\left( -\right) }$ &  Slowly varying  Rabi frequency of  forward (backward)  \\
& emission from $\vert 2\rangle\rightarrow\vert 1 \rangle$ transition.\\
$I_\mathrm{avg}^\pm\left(t \right)$ & Average temporal intensity of $\Omega^\pm$.\\
$S_\mathrm{avg}^\pm\left(\omega \right)$ & Average spectral intensity of $\Omega^\pm$.\\
$F^{+\left( -\right) }$ &  Gaussian white noise and the source term of   \\
& forward (backward) emission.\\
$\delta\left( \tau-t\right) $ &  Dirac delta function. \\
$\varphi$ & Solid angle for collecting emitted photons in the \\ 
& forward and backward direction. \\
$\tau_2$ & Lifetime of state $\vert 2 \rangle$.\\
$\Gamma$ & $=1/\tau_2$, spontaneous decay rate of state $\vert 2 \rangle$.\\
$\lambda$ &  Wavelength of $\vert 1\rangle\rightarrow\vert 2 \rangle$ transition.\\
$\rho_{ii}$ &  $=B_i B_i^*$ for state vector $\sum_{i=0}^2 B_i \vert i\rangle$, \\
& the diagonal density matrix element. \\
$\rho_{21}^{+\left( -\right) }$ &  Forward (backward) component of the coherence\\
& $B_2 B_1^*$ for state vector $\sum_{i=0}^2 B_i \vert i\rangle$.\\
$\delta_{ij}$ &  Kronecker delta symbol. \\
$n$  & Number density of particles.\\
$\sigma$ & Absorption cross section of transition $\vert 0\rangle \rightarrow\vert 2\rangle$.\\
$\sigma_r$ & $=3\lambda^2/\left( 2\pi\right) $, resonant cross section of transition $\vert 1\rangle \rightarrow\vert 2\rangle$.\\
$L$  & Length of the medium. \\
$L_g$  &   $= c \tau_p/2$, transient gain length. \\
$\alpha$ & $=n\sigma_r L$, optical depth of $\vert 1\rangle\rightarrow\vert 2 \rangle$ transition.\\
$\eta$ & $=\Gamma\alpha/\left( 2L\right) $, the light-matter coupling constant.\\
$\omega$ &  Angular frequency of $\vert 1\rangle\rightarrow\vert 2 \rangle$ transition.\\
$d$ &  Transition dipole moment of  $\vert 1\rangle\rightarrow\vert 2 \rangle$ transition.\\
$J_{p}$ & Intensity of pumping light pulse. \\
$\tau_{p}$ & Pulse duration of pumping  laser pulse.\\
$r$   & Radius of $J_{p}$ laser spot which determines the transverse\\
        & radius of the gain medium.\\
$\tau_{i}$ &  Peak position of incident pumping  laser pulse.\\
$n_{p}$ &  Number of photons per pumping  laser pulse.\\
$n_{e}$ &  Number of photons emitted by $\vert 2\rangle\rightarrow\vert 1 \rangle$ transition.\\
$Q$       & A parameter utilized to adjust the amplitude of $J_p$ in Fig.~\ref{fig4}.\\
$\tau_{R}$ & Characteristic duration of superfluorescence \cite{Friedberg1976, Schuurmans1979, Boyd1987}.\\
$\tau_{D}$ & Delay time of superfluorescence \cite{Friedberg1976, Schuurmans1979, Boyd1987}.\\
\hline
\hline
\end{tabular}
\end{table}
Figure \ref{fig1}(a) illustrates our three-level $\Lambda$-type system. A pumping light pulse $J_p$ incoherently drives transition $\vert 0\rangle \rightarrow\vert 2\rangle$, e.g., ionization, when  propagating through a one dimensional gas medium as demonstrated in Fig \ref{fig1}(b). 
The orange dots and yellow filled circles, respectively, denote particles in state $\vert 0\rangle$ and $\vert 2\rangle$. 
For simplicity, promoted particles in state $\vert 2\rangle$ subsequently experience only one decay channel $\vert 2\rangle \rightarrow\vert 1\rangle$ and emit photons in both the forward and backward direction with equal probability (red wiggled arrows). Red dashed and filled circles represent decayed particles in state $\vert 1\rangle$, and
the green Gaussian pulse depicts $J_p$.
We numerically analyse the emission behaviour for different parameters of the medium and that of $J_p$.
The Maxwell-Bloch equation \cite{Arecchi1970, MacGillivray1976, Polder1979, Schuurmans1979, Yuan2012, Weninger2014} with forward-backward decomposition \cite{Lin2009, Liao2014, Su2016} is used to describe the dynamics including the incoherent and longitudinal pumping:
\begin{eqnarray}
\partial_t \rho_{00} &=&-\sigma J_p \rho_{00}, \label{eq1}\\
\partial_t \rho_{11} &=&\Gamma\rho_{22}-\frac{i}{2}\left( \Omega^+\rho_{21}^{+*}-\Omega^{+*}\rho_{21}^+\right)\nonumber\\
&-&\frac{i}{2}\left( \Omega^-\rho_{21}^{-*}-\Omega^{-*}\rho_{21}^-\right), \label{eq2}\\
\partial_t \rho_{22} &=&\sigma J_p \rho_{00}-\Gamma\rho_{22}
+\frac{i}{2}\left( \Omega^+\rho_{21}^{+*}-\Omega^{+*}\rho_{21}^+\right)\nonumber \\
&+&\frac{i}{2}\left( \Omega^-\rho_{21}^{-*}-\Omega^{-*}\rho_{21}^-\right),  \label{eq3}
\end{eqnarray}
\begin{eqnarray}
\partial_t \rho_{21}^+ &=& -\frac{\Gamma}{2}\rho_{21}^+ -\frac{i}{2}\left(\rho_{22}-\rho_{11} \right)\Omega^+ +F^+ ,\\
\partial_t \rho_{21}^- &=& -\frac{\Gamma}{2}\rho_{21}^- -\frac{i}{2}\left(\rho_{22}-\rho_{11} \right)\Omega^- +F^- ,\\
%
\frac{1}{c}\partial_t J_p &+& \partial_z J_p =-n\rho_{00}  \sigma J_p,\label{eq6}\\
\frac{1}{c}\partial_t \Omega^+ &+& \partial_z \Omega^+ =i\eta\rho_{21}^+,\label{eq7}\\
\frac{1}{c}\partial_t \Omega^- &-& \partial_z \Omega^- =i\eta\rho_{21}^-; 
\end{eqnarray}
together with initial and boundary conditions
\begin{eqnarray}
\rho_{ij}\left( 0, z\right) & = &\delta_{i0}\delta_{j0}, \\
\rho_{21}^{\pm}\left( 0, z\right) & = & 0, \\
J_p\left( 0, z\right) & = & 0, \\
\Omega^{\pm}\left( 0, z\right) & = & 0, \\
J_p\left( t, 0\right) & = & \frac{n_p}{\pi^{3/2}r^2 \tau_p} Exp\left[ -\left( \frac{t-\tau_i}{\tau_p}\right)^2 \right] , \label{eq13}\\
\Omega^{+}\left( t, 0\right) & = & 0, \\
\Omega^{-}\left( t, L\right) & = & 0.
\end{eqnarray}
The emission process starts from  Gaussian white noise $F^{\pm}$  obeying the delta correlation function \cite{Weninger2014, Haake1979, Haake1979b, Polder1979, Gross1982}
\begin{equation}
\langle F^{\pm}(\tau)F^{\pm *}(t)\rangle =\frac{\varphi\rho_{22}\Gamma^2\omega^2}{24 n \pi^2 c^3}\delta(\tau-t), \label{eq16}
\end{equation}
where 
\begin{equation}
\varphi = \int_0^{2\pi}d\phi\int_0^{\tan^{-1}\left( \frac{r}{L}\right)} \sin\theta d\theta=2\pi-\frac{2\pi}{\sqrt{1+\frac{r^2}{L^2}}}. \nonumber
\end{equation}
Based on an experimental fact \cite{Vrehen1981} and \cite{Gross1982}, we use
\begin{equation}
\langle F^{\pm}(\tau)F^{\mp *}(t)\rangle =0. \nonumber
\end{equation}
All  notation in above equations is listed and explained in Table~\ref{table1}, and the  quantities used for each  figure in what follows are listed in Table~\ref{table2}. Fig.~\ref{fig1}(c) illustrates the solid angle $\varphi$ used in Eq.~(\ref{eq16}) \cite{Arecchi1970}. $\varphi$ is determined by the geometry of the gain medium, i.e., an ensemble of atoms longitudinally  pumped to state $\vert 2\rangle$ by a $J_p$ pulse. $\varphi$ is therefore affected by the length of the medium $L$ and the radius $r$ of the $J_p$ laser spot. Those photons randomly emitted within $\varphi$ will interact with most of excited atoms and lead to, e.g., stimulated emission.
The value from the complete form of $\varphi$ is used in all of our numerical calculations, and
$\varphi$ can be simplified to an intuitive value $\pi r^2/L^2$ for typical theoretical studies since $r\ll L$ is implicitly assumed in the 1D model. However, in realistic systems, diffraction (3D effects) can limit the effective solid angle to $\lambda^2/r^2$, and can change the simulation results. This problem is associated with Fresnel number $\mathbf{F}$ \cite{Gross1982} and will be discussed in Sec.~\ref{sec:nr}.
The ensemble average of temporal intensity $\langle I^{\pm} \left(t \right) \rangle$ is defined as 
\begin{eqnarray}
\langle I^{+} \left(t \right) \rangle &=& \frac{1}{N_e}\sum_{n=1}^{N_e} \vert\Omega_n^+\left( t,L\right) \vert^{2}, \nonumber\\
\langle I^{-} \left(t \right) \rangle &=& \frac{1}{N_e}\sum_{n=1}^{N_e} \vert\Omega_n^-\left( t,0\right) \vert^{2}.\label{eq17}
\end{eqnarray}
and the ensemble average of spectral intensity $\langle S^\pm\left(\omega \right)\rangle$ is defined as 
\begin{eqnarray}
\langle S^+\left(\omega \right)\rangle &=& \frac{1}{N_e}\sum_{n=1}^{N_e} \vert\int_{-\infty}^{\infty}\Omega_n^+\left( t,L\right) e^{i\omega t}dt\vert^{2}, \nonumber\\
\langle S^-\left(\omega \right)\rangle &=& \frac{1}{N_e}\sum_{n=1}^{N_e} \vert\int_{-\infty}^{\infty}\Omega_n^-\left( t,0\right) e^{i\omega t}dt\vert^{2}. \label{eq18}
\end{eqnarray}
Here $\Omega_n^\pm$ is the output $\Omega^\pm$ of $n$th simulation, and $N_e = 1000$ is the sample size. Our sample size of 1000 is chosen by a series of numerical tests showing that convergence occurs in a range of  $N_e = 500-900$, depending on parameters.
\begin{widetext}
\begin{table*}[t]
\centering
\caption{\label{table2}
Quantities used in each figure.
The common quantities are $c=3\times 10^8$m/s, $\varepsilon_0 = 8.85\times 10^{-12}$ F/m, $\sigma = 3.336 \times 10^{-23} m^{2}$, $\sigma_r = 6.4 \times 10^{-18} m^{2}$, $\omega=1.29 \times 10^{6}$rad$\cdot$THz, $\lambda=1.46$nm and $d=3.33 \times 10^{-31}$C$\cdot$m. Symbol $V$ stands for variable.
}
\begin{tabular}{llllllllll}
\hline
\hline
Notation & Fig.~\ref{fig3}(a) &  Fig.~\ref{fig3}(b) &  Fig.~\ref{fig4} &  Fig.~\ref{fig5}\&\ref{fig2}\&\ref{fig8} &  Fig.~\ref{fig6}\&\ref{fig9} &  Fig.~\ref{fig7}  &  Fig.~\ref{fig10}\\
\hline
$\varphi$($\mu$rad) & 12.56 & 12.56 & 12.56 & $V$ & $V$ & $V$  & 12.56 \\
$\tau_2$(ps) & 1 & 100 & 0.01 & 1 & 100 & 0.01  & 1 \\
$\Gamma$(THz) & 1 & 0.01 & 100 & 1 & 0.01 & 100  & 1 \\
$n$(cm$^{-3}$)  & $3 \times 10^{16}$ & $3 \times 10^{16}$ & $3 \times 10^{16}$ & $V$ & $V$ & $V$ &  $5 \times 10^{17}$ \\
$L$(mm)  & 1 & 1 & 1 & $V$ & $V$ & $V$ & 0.16 \\
$\alpha$ & 192 & 192 & 192 & $V$ & $V$ & $V$ & 512 \\
$\eta$(THz/mm) & 96 & 0.96 & 9600 & $V$ & $V$ & $V$ & 1600 \\
$\tau_{p}$(fs) & 60 & 60 & $V$ & 60 & 60 & 60 & 60 \\
$r$($\mu$m)   & 2 & 2 & 2 & 2 & 2 & 2 & $V$ \\
$\tau_{i}$(ps) & 0.3 & 30 & 0.3 & 0.24 & 0.24 & 0.24 & 0.24 \\
$n_{p}$ & $30 \times 10^{12}$ & $30 \times 10^{12}$ & $V$ & $30 \times 10^{12}$ & $30 \times 10^{12}$ & $30 \times 10^{12}$ & $V$ \\
\hline
\hline
\end{tabular}
\end{table*}
\end{widetext}

\begin{figure}[b]
\vspace{-0.4cm}
  \includegraphics[width=0.45\textwidth]{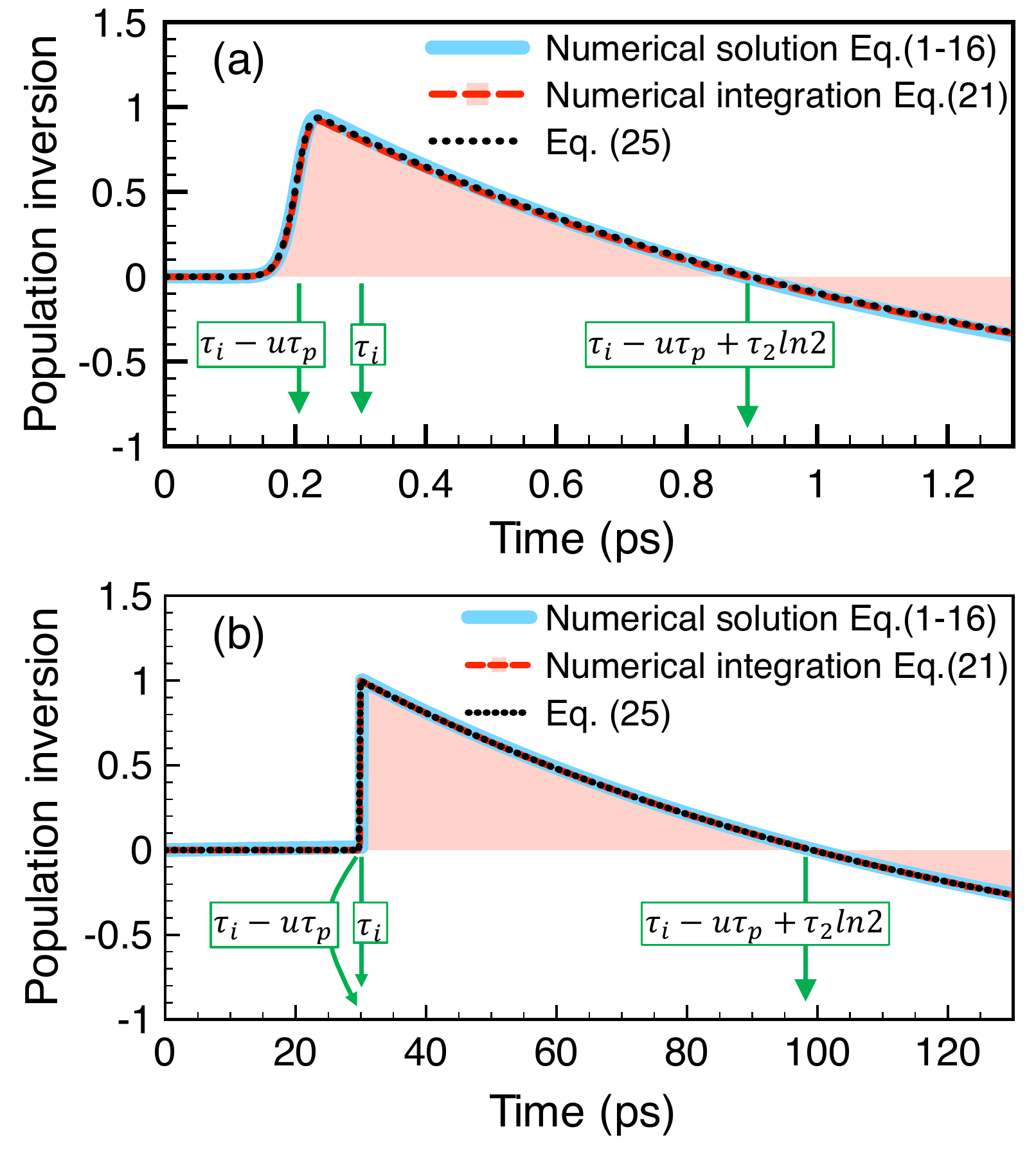}
  \caption{\label{fig3}
(Color online) Population inversion $I(t, z = 0)$ for (a) $(\tau_2, \tau_i)=(1\mathrm{ps}, 0.3\mathrm{ps})$ and (b) $(\tau_2, \tau_i)=(100\mathrm{ps}, 30\mathrm{ps})$. Thick-blue-solid, red-dashed-filled and black-dotted lines depict $\rho_{22}(t,0)-\rho_{11}(t,0)$ from 
the numerical solution of Eq.~(\ref{eq1}-\ref{eq16}), numerical integration of Eq.~(\ref{eq21}) and Eq.~(\ref{eq25}), respectively. Three downward green arrows chronologically indicate three key temporal instants, namely, $\tau_i-u\tau_p$, $\tau_i$ and $\tau_i-u\tau_p+\tau_2\ln 2$. The gain duration of $I(t, z)>0$ is about $\tau_2\ln 2$. Other parameters are $(r, n_p, \tau_p, \sigma)=(2\mu\mathrm{m}, 30\times 10^{12}, 60\mathrm{fs}, 3.336\times 10^{-23} \mathrm{m^2})$.
  }
\end{figure}
\begin{figure}[b]
\vspace{-0.4cm}
  \includegraphics[width=0.45\textwidth]{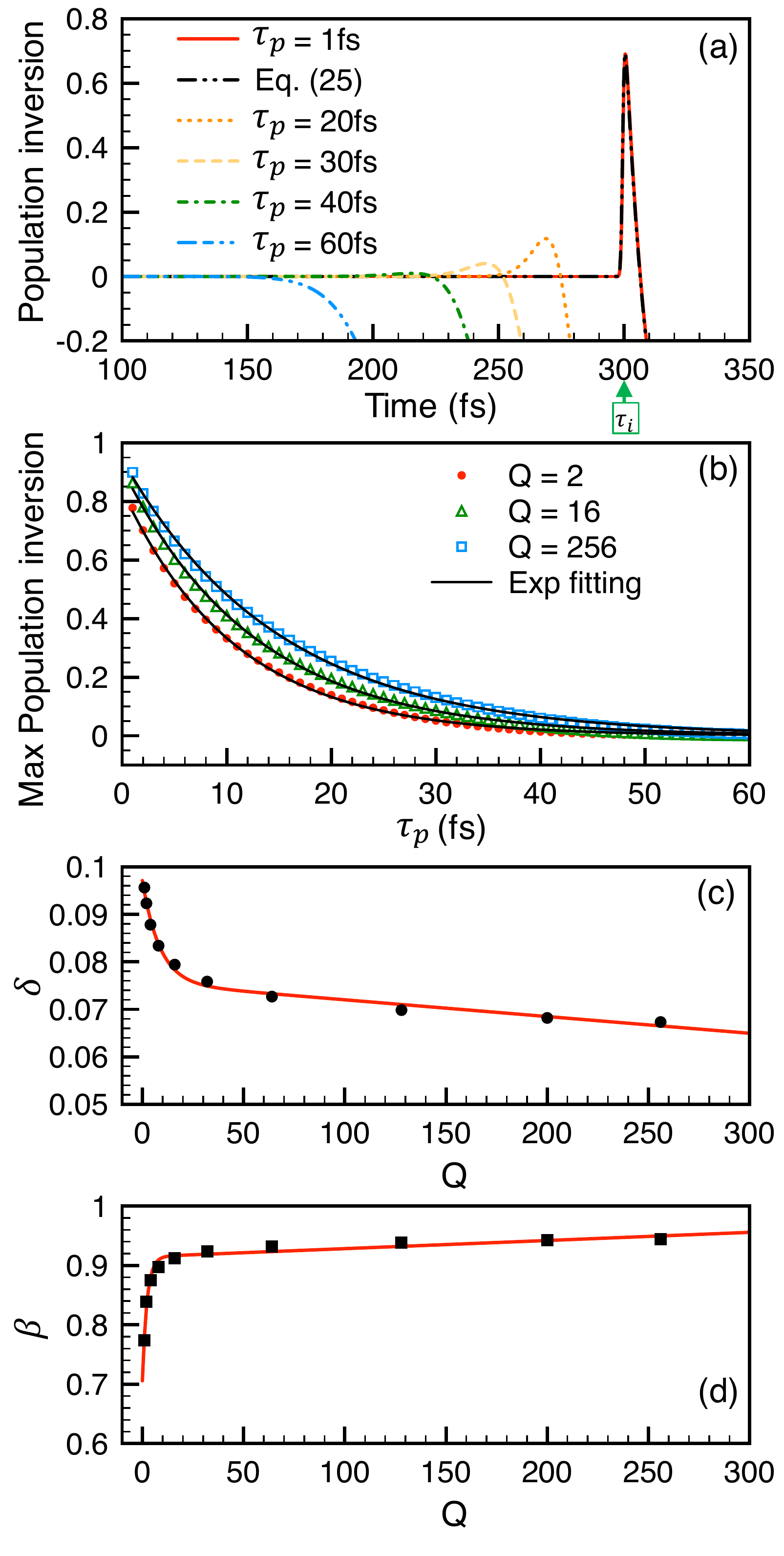}
  \caption{\label{fig4}
(Color online) (a) population inversion $I(t, z = 0)$ for $(\tau_2, \tau_i) = (10\mathrm{fs}, 300\mathrm{fs})$ and a range of $J_p$ pulse duration based on XFEL parameter, namely, $\tau_p=1$fs (red solid line), $\tau_p=20$fs (orange dotted line), $\tau_p=30$fs (yellow dashed line), $\tau_p=40$fs (green dashed-dotted line) and $\tau_p=60$fs (blue dashed-dashed-dotted-dotted line).  Black dashed-dotted-dotted line is Eq.~(\ref{eq25}). Other parameters are $(n_p, \tau_p, r, \sigma)=(Q T_p \times 10^{12}, T_p \mathrm{fs}, 2 \mu\mathrm{m} , 3.336\times 10^{-23} \mathrm{m^2})$ and $Q=1$. Upward green arrow  indicates $\tau_i$.
(b) the max population inversion as a function of $T_p$ for $Q=2$ (red dots), $Q=16$ (green triangles) and $Q=256$ (blue squares). Black solid lines are $\beta e^{-\delta T_p}$ fittings.
(c) $Q$-dependent $\delta$. (d) $Q$-dependent $\beta$.
  }
\end{figure}
\section{Analytical solutions}\label{sec:as}
Given the XFEL pulse duration $\leq$100 fs and possible wide range of $\tau_2$ for different systems, it is necessary to analyse the influence of temporal structure of pumping pulse on the production of population inversion.
In this section we analyse the pumping process in the region of $\vert \Omega^{\pm} \vert \ll \Gamma  < J_p \sigma$, namely, pumping rate is greater than excited state decay rate and photon emission rate, which  allows for the production of population inversion.
By first using $\rho_{00}\left( t, z\right)=1$, the solution of Eq.~(\ref{eq6}) reads
\begin{equation}
J_p(t,z) = \frac{n_p}{\pi^{3/2}r^2 \tau_p} \mathrm{Exp}\left[ -n\sigma z-\left( \frac{t-\tau_i - \frac{z}{c}}{\tau_p}\right)^2 \right].
\end{equation}
In the parameter region of Fig.~\ref{fig3} and Fig.~\ref{fig4},  $\mathrm{Exp}\left( -n\sigma L\right)>0.98$, one can therefore neglect the attenuation of $J_p$ for both cases. The solution of Eq.~(\ref{eq1}) is then given by
\begin{equation}\label{eq20}
\rho_{00}(t, z)\approx \mathrm{Exp} \left\lbrace  -\frac{n_p \sigma }{2\pi r^2}\left[  1 +\mathrm{erf}\left( \frac{t-\tau_i - \frac{z}{c}}{\tau_p}\right)  \right]   \right\rbrace.
\end{equation}
The dynamics of $\rho_{22}$ obey $\partial_t \rho_{22} =\sigma J_p \rho_{00}-\Gamma\rho_{22}$ whose solution reads
\begin{eqnarray}
&&\rho_{22}(t) \approx  \frac{n_p \sigma}{\pi^{3/2}r^2 \tau_p} \mathrm{Exp}\left( -\Gamma t -\frac{n_p \sigma }{2\pi r^2}\right) \nonumber\\
&\times & \int_0^{t}\mathrm{Exp}\left[ \Gamma s-\frac{n_p \sigma }{2\pi r^2} \mathrm{erf}\left( \frac{s-\tau_i}{\tau_p}\right)-\left( \frac{s-\tau_i}{\tau_p}\right)^2    \right] ds. \label{eq21}\nonumber\\
\end{eqnarray}
Invoking the conservation of population $\sum_{i=1}^{3}\rho_{ii}(t,z) =1$ one gets $\rho_{11}$.
A careful comparison confirms that  Eq.~(\ref{eq21}) is equivalent to the numerical solution of complete Eqs.~(\ref{eq1}-\ref{eq16}). 
When $\tau_p<\tau_2$, we can obtain  approximate solutions
\begin{eqnarray}
\rho_{22}(t,z) &\approx &  \left[1- \rho_{00}(t)\right]e^{-\Gamma \left( t-\tau_i-\frac{z}{c}+u\tau_p\right) }, \\
\rho_{11}(t,z) &\approx &  \left[1- \rho_{00}(t)\right]\left[ 1-e^{-\Gamma \left( t-\tau_i-\frac{z}{c}+u\tau_p\right) }\right],\\
u & = & 2 +
\frac{e\pi^{3/2}r^2}{n_p\sigma}
\left( 1-\sqrt{1+\frac{4 n_p\sigma}{e\pi^{3/2}r^2}}\right) ,
\end{eqnarray}
where  $u\tau_p$ corresponds to $\partial^2_t \rho_{00}(t)\vert_{t=u\tau_p}=0$  when the maximum consumption rate of $\rho_{00}$ by $J_p$ occurs, and $e$ is Euler's number. 
One can also obtain the approximate population inversion $I(t,z)=\rho_{22}(t,z) -\rho_{11}(t,z)$
\begin{equation}\label{eq25}
I(t,z)\approx  \left[ 1- \rho_{00}(t, z)\right]\left[ 2 e^{-\Gamma \left( t-\tau_i -\frac{z}{c} + u\tau_p\right) } -1\right]. 
\end{equation}
Figure~\ref{fig3} demonstrates the comparison between the  numerical solution of complete Eqs.~(\ref{eq1}-\ref{eq16}), the numerical integration of Eq.~(\ref{eq21}) and the analytical solution of  Eq.~(\ref{eq25}) for $\tau_2=1$ps and $\tau_2=100$ps. This plainly demonstrates the consistency between the three methods.
When the time scale of $\tau_2$ is  small and approaching $\tau_p$, the effect caused by  the leading edge of the $J_p$ pulse becomes significant, i.e., the shift $u\tau_p$ of around 0.2ps in Fig.~\ref{fig3}(a). The peak value of  $I(t,z)$ actually  happens at about $\tau_i+\frac{z}{c}-u\tau_p$ instead of $\tau_i+\frac{z}{c}$. The temporal shift of $u\tau_p$, namely, the spacing between the first two downward green arrows, reveals that the subsequent emissions may overtake $J_p$. The excited atoms subsequently experience spontaneous decay, and give a similar duration of $\tau_2\ln 2$, i.e., the spacing between first and third downward green arrows, for positive $I(t,z)$.
This gain domain, e.g., $[0.2\mathrm{ps},0.9\mathrm{ps}]$ in Fig.~\ref{fig3}(a) and $[30\mathrm{ps},98\mathrm{ps}]$ in Fig.~\ref{fig3}(b), is the  energy reservoir for ASE while outside the region $\Omega^{\pm}$ will be reabsorbed by atoms in the ground state. On can accordingly define the transient gain length $L_g = c \tau_p/2$ in the medium. The analysis of $I(t,z)$ is important for the understanding of the ASE gain curve, which will be performed in what follows.

In order to investigate the gain process of $\Omega^+$, we further approximate
\begin{equation}
I(t,z)\approx  \left[ 2 e^{-\Gamma \left( t-\tau_i -\frac{z}{c} + u\tau_p\right) } -1\right] \Theta\left( t- \tau_i-\frac{z}{c}+u\tau_p \right), \label{eq26}
\end{equation}
where $\Theta$ is the Heaviside unit step function, and
solve the simplified equations
$\partial_t \rho_{21}^+ = -\frac{i}{2}P I\Omega^+$ and $ \partial_z \Omega^+ =i\eta\rho_{21}^+$.
The integral of the former leads to $\rho_{21}^+ (t, z)= -\frac{i}{2}P\int_0^{t} \Omega^+ (\tau, z) I(\tau,z) d\tau$ which is then substituted into the latter.
We arrive at
\begin{eqnarray}
\partial_z\Omega^+ (t, z) 
& = &
\frac{P\Gamma\alpha}{4 L} \int_0^{t} \Omega^+ (\tau, z) I(\tau,z) d\tau \nonumber\\
& \approx &
\Omega^+ (t, z) \frac{P\Gamma\alpha}{4 L} \int_0^{t}  I(\tau,z) d\tau. \label{eq27}
\end{eqnarray}
Neglecting $ \partial_t \Omega^+$ in the wave equation  is justified when Eq.~(\ref{eq26}) goes to the retarded
time frame, namely, as a function of $t-\frac{z}{c}$ \cite{Bonifacio1975, MacGillivray1976, Rai1992}.
The additional factor $P$ indicates the weighting of the forward emission. In the early stage, i.e., $\Gamma\gg \vert\Omega^+\vert$, spontaneous emission dominates, and so $P = 1/2$ is expected in the simplified equations because an atom has the equal chance for the forward/backward spontaneous emission.
We further neglect the temporal structure of $\Omega^+ (t, z)$ and extract it from the integrand of  Eq.~(\ref{eq27}).
The peak value of output $\Omega^+$ pulse in the end of the medium $(z=L)$ reads
\begin{eqnarray}
\Omega^+ (L) 
&\approx &
\Omega^+_0 
\mathrm{Exp}\left\lbrace \frac{P\Gamma\alpha}{4 L}\int_0^L \int_{\tau_l}^{\tau_u} I(\tau,z) d\tau dz\right\rbrace\nonumber\\
& = &
\Omega^+_0 
e^{\xi\alpha}.
\end{eqnarray}
Here $\Omega^+_0$ is given by Gaussian white noise, and $\tau_l =\tau_i+\frac{z}{v}$, $\tau_u=\tau_i+\frac{z}{c}-u\tau_p +\tau_2\ln 2$.
Because $\Omega^+$ is always behind $J_p$ and must stay in the gain interval $\left[ \tau_l, \tau_u \right] $, where the population inversion $I>0$, as demonstrated in Fig.~\ref{fig3}. 
The gain factor $2\xi$ is then determined to be
\begin{widetext}
\begin{equation}\label{eq30}
2\xi=\frac{P}{4}\left\lbrace \frac{L\Gamma}{v}-\frac{L\Gamma}{c}+2 u \tau_p\Gamma - 2 -\ln 4 
-\frac{4 c v }{\left( c-v\right)L\Gamma }e^{-u  \tau_p\Gamma}\left[e^{\left( \frac{1}{c}-\frac{1}{v}\right) L\Gamma }-1 \right]\right\rbrace.
\end{equation}
\end{widetext}
The upper bound of $2\xi$ in Eq.~(\ref{eq30}) occurs when $v$ approaches $c$, namely, 
\begin{equation}\label{eq31}
\lim_{v\to c} 2\xi=\frac{P}{2}\left( 2 e^{-u\tau_p\Gamma}+u\tau_p\Gamma-1-\ln 2\right),
\end{equation}
which results in the maximum gain exponent. 
One can deduce the range of the group velocity $v$ of the emitted light pulse as follows. 
Given the fact that the $\Omega^+$ pulse propagates with group velocity $v$ all the way behind $J_p$ and so $v < c$.
Since the emitted light must stay in the gain domain illustrated in Fig.~\ref{fig3} when $I\left( t,z\right) > 0$, otherwise it will be re-absorbed, we have the gain condition $\tau_i+\frac{z}{c}\leq \tau_l\leq\tau_u$. This  indicates that $J_p$ produces population inversion at $z$ during $\left[ \tau_i+\frac{z}{c}, \tau_u\right]$ where we neglect the small correction of $u\tau_p$ in the lower bound, but $\Omega^+$ passes through $z$ at $\tau_l$ which must be within the above gain interval. 
The inverse of the gain condition leads to the range of $v$
\begin{equation}
c\geq v \geq \frac{c L \Gamma}{L\Gamma+c\ln 2-c\Gamma u\tau_p}. \label{eq29}
\end{equation}
We analyse the trajectory of the emitted pulse and observe that the group velocity $v$ is initially smaller than $c$ and gradually approaches $c$ when the optical  ringing effect occurs. While the acceleration mechanism remains an interesting theoretical topic to be studied, one can still use Eq.~(\ref{eq31}) to estimate the upper bound of ASE gain exponent.

\begin{figure*}[t]
\vspace{-0.4cm}
  \includegraphics[width=1\textwidth]{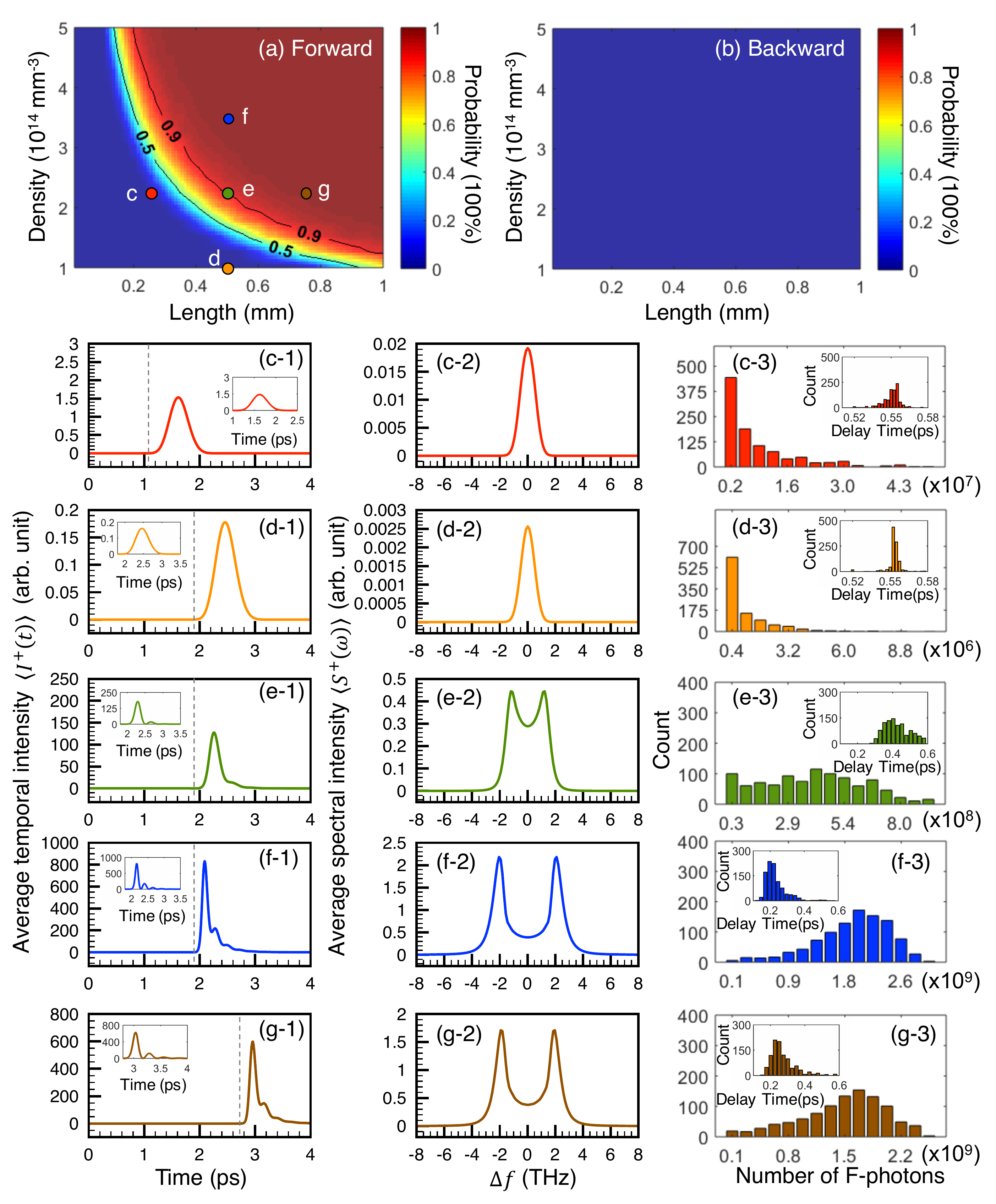}
  \caption{\label{fig5}
(Color online) Probability of (a) forward $\int_{-\infty}^\infty\vert\Omega^{+}\left( t, L\right)  \vert dt \geq \frac{\pi}{2}$ and that of (b) backward $\int_{-\infty}^\infty\vert\Omega^{-}\left( t, 0\right)  \vert dt \geq \frac{\pi}{2}$  as a function of ($L$, $n$)  among 1000 realizations with Gaussian random noise for ($r$, $n_p$, $\tau_p$, $\tau_2$)=(2$\mu$m, 30$\times 10^{12}$, 60fs, 1ps). 
Five data sets, marked as \textbf{c} ($L$, $n$) = (0.25mm, 2.25$\times 10^{14}$mm$^{-3}$), \textbf{d} (0.5mm, 1$\times 10^{14}$mm$^{-3}$), \textbf{e} (0.5mm, 2.25$\times 10^{14}$mm$^{-3}$), \textbf{f} (0.5mm, 3.5$\times 10^{14}$mm$^{-3}$) and \textbf{g} (0.75mm, 2.25$\times 10^{14}$mm$^{-3}$) in (a), are picked up to show the transition from amplified spontaneous emission to superfluorescence of $\Omega^+$. The corresponding average temporal intensity, average spectral intensity and photon number histogram of each chosen point are respectively illustrated by ($\textbf{x}-1$), ($\textbf{x}-2$) and ($\textbf{x}-3$), where $\textbf{x}\in \left\lbrace \textbf{c}, \textbf{b}, \textbf{e}, \textbf{f}, \textbf{g} \right\rbrace $.
Gray dashed lines in (\textbf{c}-1)-(\textbf{g}-1) indicate instants when $J_{p}$ leaves the medium.
Insets in (\textbf{c}-1)-(\textbf{g}-1)
and (\textbf{c}-3)-(\textbf{g}-3)
demonstrate the typical $\vert\Omega^{+}\left( t, L\right)\vert^2$ of a single realization and delay time histogram of forward emission at each chosen set of parameters, respectively. 
  }
\end{figure*}
%
\begin{figure*}[t]
\vspace{-0.4cm}
  \includegraphics[width=1\textwidth]{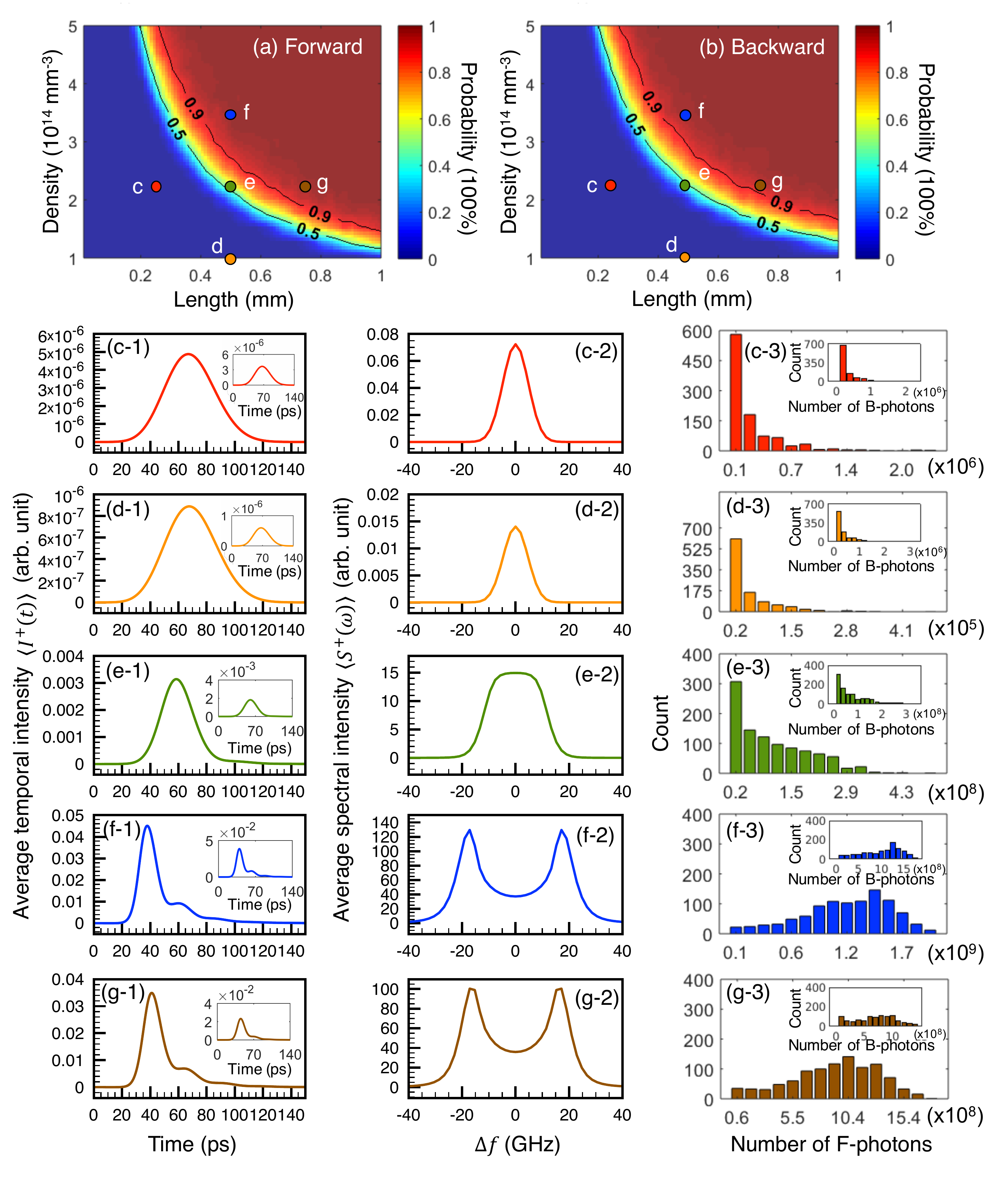}
  \caption{\label{fig6}
(Color online) Probability of (a) forward $\int_{-\infty}^\infty\vert\Omega^{+}\left( t, L\right)  \vert dt \geq \frac{\pi}{2}$ and that of (b) backward $\int_{-\infty}^\infty\vert\Omega^{-}\left( t, 0\right)  \vert dt \geq \frac{\pi}{2}$ as a function of ($L$, $n$) among 1000 realizations with Gaussian random noise for ($r$, $n_p$, $\tau_p$, $\tau_2$)=(2$\mu$m, 30$\times 10^{12}$, 60fs, 100ps). 
%
Five data sets, marked as \textbf{c} ($L$, $n$) = (0.25mm, 2.25$\times 10^{14}$mm$^{-3}$), \textbf{d} (0.5mm, 1$\times 10^{14}$mm$^{-3}$), \textbf{e} (0.5mm, 2.25$\times 10^{14}$mm$^{-3}$), \textbf{f} (0.5mm, 3.5$\times 10^{14}$mm$^{-3}$) and \textbf{g} (0.75mm, 2.25$\times 10^{14}$mm$^{-3}$) in (a) and (b), are chosen for showing the transition from amplified spontaneous emission to superfluorescence.
The corresponding average temporal intensity, average spectral intensity and photon  number histogram of the forward emission at each chosen point are respectively illustrated by ($\textbf{x}-1$), ($\textbf{x}-2$) and ($\textbf{x}-3$), where $\textbf{x}\in \left\lbrace \textbf{c}, \textbf{b}, \textbf{e}, \textbf{f}, \textbf{g} \right\rbrace $.
The corresponding average temporal intensity and photon  number histogram of the backward emission are illustrated in the
insets of (\textbf{c}-1)-(\textbf{g}-1)
and those of (\textbf{c}-3)-(\textbf{g}-3), respectively.
  }
\end{figure*}

In view of the time structure of XFEL \cite{EuropeanXFEL} and the recent XFEL-pumped laser \cite{Rohringer2012}, we investigate the  pumping process with a variety of XFEL parameters \cite{EuropeanXFEL}.
Fig.~\ref{fig4}(a) illustrates the numerical solution of $I(t,0)$ for $\tau_2=10$fs with a variety of $\tau_p$. 
We use $(n_p, \tau_p, \tau_i, r)=(Q T_p \times 10^{12}, T_p \mathrm{fs}, 300\mathrm{fs}, 2 \mu\mathrm{m})$ and $Q=1$ to
fix the amplitude of $J_p$ at a constant $4.5 \times 10^{37}$ s$^{-1}$m$^{-2}$ when varying $T_p$. The red solid line depicts the case for $\tau_p=1$fs, i.e., $T_p=1$ in which $J_p$ is short enough to generate population inversion between state $\vert 2 \rangle$ and $\vert 1 \rangle$ with a peak value of 0.69. 
As one can see, the analytical solution Eq.~(\ref{eq25}) (black dashed-dotted-dotted line) also matches the numerical integration of Eq.~(\ref{eq21}) (red solid line) as $\tau_p <\tau_2$.
When $\tau_p\geq\tau_2$, the analysis of Eq.~(\ref{eq21}) becomes a complicated problem, and 
the  competition between the $\sigma J_p$ and $\Gamma$ terms in Eq.~(\ref{eq3}) causes $I(t,z)$ to exhibit more complicated behaviour. 
For long pumping pulses of $\tau_p=20$fs (orange dotted line), $\tau_p=30$fs (yellow dashed line), $\tau_p=40$fs (green dashed-dotted line) and $\tau_p=60$fs (blue dashed-dashed-dotted-dotted line), the $\tau_p$-dependent reduction of population inversion is observed. Such a reduction is caused by the consumption of $\rho_{00}$ by the leading edge of $J_p$ whose pumping power is too weak to compete against the fast decay of state $\vert 2 \rangle$. The leading edge of $J_p$ and the decay rate $\Gamma$ turn the medium transparent 
before the arrival of the peak of $J_p$. 
Fig.~\ref{fig4}(b-c) show that the increase of the amplitude of $J_p$ slightly eases the reduction of population inversion due to the leading edge of $J_p$. In Fig.~\ref{fig4}(b) we use $Q=2$ (red dots), $Q=16$ (green triangles) and $Q=256$ (blue squares) to demonstrate the effect of increasing the $J_p$ amplitude
on the maximum population inversion as a function of  $T_p$. 
The black solid lines are $\beta e^{-\delta T_p}$ fittings. 
The $Q$-dependent  $\delta$ and $\beta$  are respectively depicted in Fig.~\ref{fig4}(c) and Fig.~\ref{fig4}(d).
As one can see in Fig.~\ref{fig4}(c) that, in the $Q$ domain of $\left[ 1, 300\right] $, $\delta$ ranges between 0.07 to 0.1 corresponding to a variety of $\tau_p$ between 10fs and 15fs. 
When $\tau_p$ is shorter than this range, transient population inversion can be built up and this results in a noticeable gain of $\Omega^{\pm}$.
For $\tau_p > 40$fs population inversion is suppressed and the gain of $\Omega^{\pm}$ becomes negligible.
Consequently, the temporal structure of a pumping pulse plays a more critical role than  $J_p$ amplitude does when $\tau_p$ and $\tau_2$ are in a similar time scale. 
When $\tau_p$ is too long, the pumping capability of a $J_p$ pulse will degrade.
The rapid decrease of $\delta$ in Fig.~\ref{fig4}(c) and the quick increase of $\beta$ in Fig.~\ref{fig4}(d) for $Q<$ 20 suggest that there is an efficient choice of $J_p$ amplitude to optimize the generation of population inversion. It is worth mentioning that the red solid fitting curves in Fig.~\ref{fig4}(c \& d) are in the form of $a+b e^{-f Q}+g Q$, where $a$, $b$, $f$ and $g$ are some fitting constants, which may provide useful information for future analytical study of $I(t,z)$.

\section{Numerical Results}\label{sec:nr}
Here we turn to the discussion of the numerical solution of Eq.~(\ref{eq1}-\ref{eq16}).
Fig.~\ref{fig5} demonstrates the results of ($r$, $n_p$, $\tau_p$, $\tau_2$) = (2$\mu$m, 30$\times 10^{12}$, 60fs, 1ps) and  Fig.~\ref{fig6} shows that of ($r$, $n_p$, $\tau_p$, $\tau_2$) = (2$\mu$m, 30$\times 10^{12}$, 60fs, 100ps).
Fig.~\ref{fig5} (a) and Fig.~\ref{fig6} (a)  depict the probability  of forward emission $\int_{-\infty}^\infty\vert\Omega^+\left( t, L \right) \vert dt\geq\frac{\pi}{2}$  among 1000 realizations of simulation at each point of ($L$, $n$). 
Fig.~\ref{fig5} (b) and Fig.~\ref{fig6} (b) depict that of backward emission $\int_{-\infty}^\infty\vert\Omega^-\left( t, 0\right)  \vert dt\geq\frac{\pi}{2}$.
In the forward emission for both $\tau_2=1$ps and $\tau_2=100$ps, the growth of the area of $\Omega^+$ and its gain are observed when either the length of the medium $L$ or the density  of the medium $n$ increases. However, for  backward emission, the similar tendency only occurs when  $\tau_2=100$ps but is absent for $\tau_2=1$ps.
Fig.~\ref{fig5} (b)  reveals that no matter how one changes the parameters of a gain medium the backward pulse area  $\int_{-\infty}^\infty\vert\Omega^-\left( t, 0\right)  \vert dt$ is always negligible. This forward-backward asymmetry will be discussed in Sec.~\ref{sec:fba}.

\begin{figure}[b]
\vspace{-0.4cm}
  \includegraphics[width=0.45\textwidth]{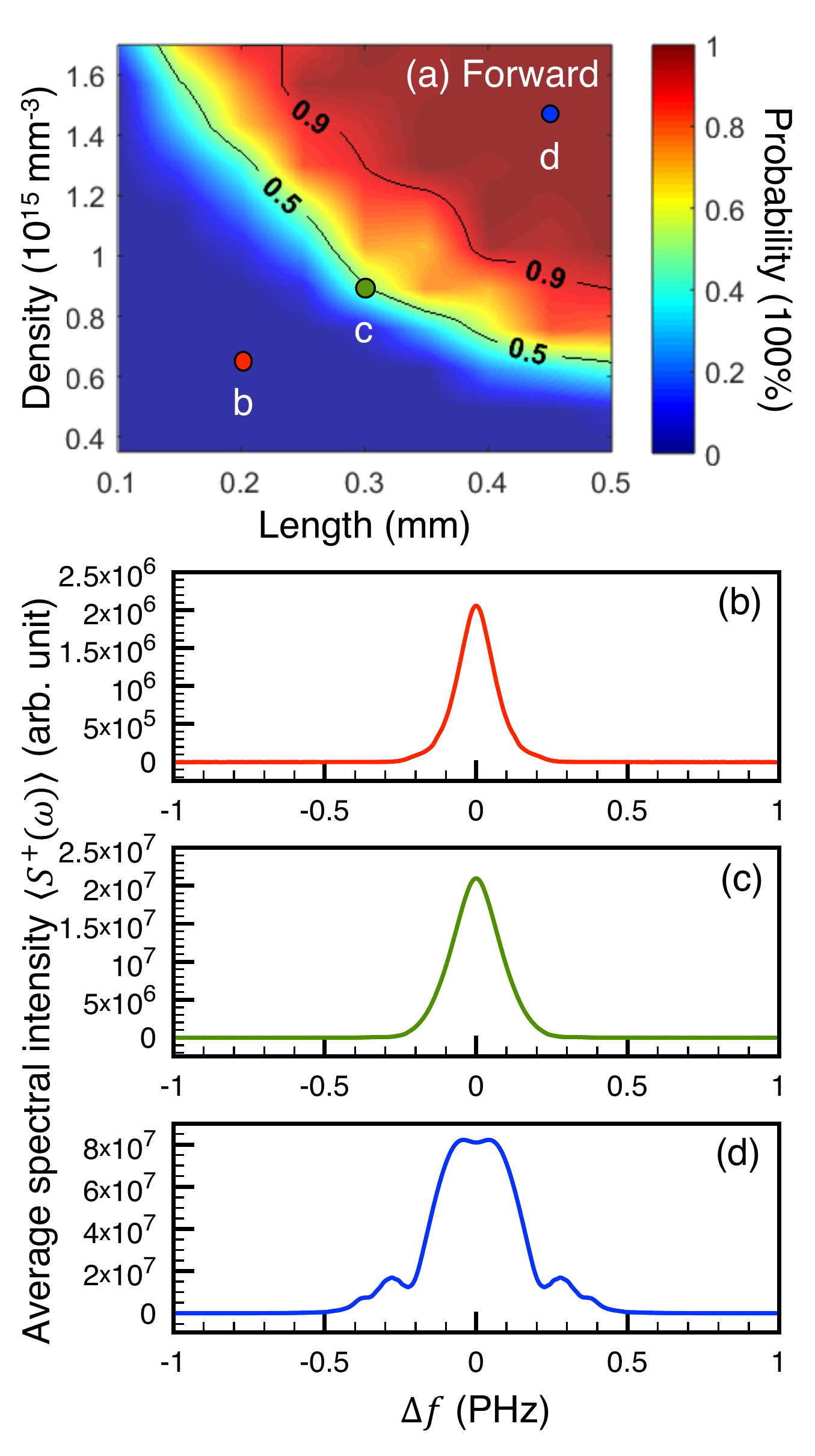}
  \caption{\label{fig7}
(Color online) (a) Probability of $\int_{-\infty}^\infty\vert\Omega^{+}\left( t, L\right)  \vert dt \geq \frac{\pi}{2}$ occurs as a function of ($L$, $n$) in forward direction among 1000 realizations with Gaussian random noise for ($r$, $n_p$, $\tau_p$, $\tau_2$)=(2$\mu$m, $10^{14}$, 24fs, 10fs). $\alpha\simeq 	1760$ for probability of 50\%.
The average spectral intensity of five data sets, marked as \textbf{b} ($L$, $n$) = (0.2mm, 6.5$\times 10^{14}$mm$^{-3}$), \textbf{c} (0.3mm, 9.5$\times 10^{14}$mm$^{-3}$), \textbf{d} (0.44mm, 1.48$\times 10^{15}$mm$^{-3}$), are chosen for showing the transition from amplified spontaneous emission to superfluorescence of $\Omega^+$.
  }
\end{figure}

\begin{figure}[b]
\vspace{-0.4cm}
  \includegraphics[width=0.48\textwidth]{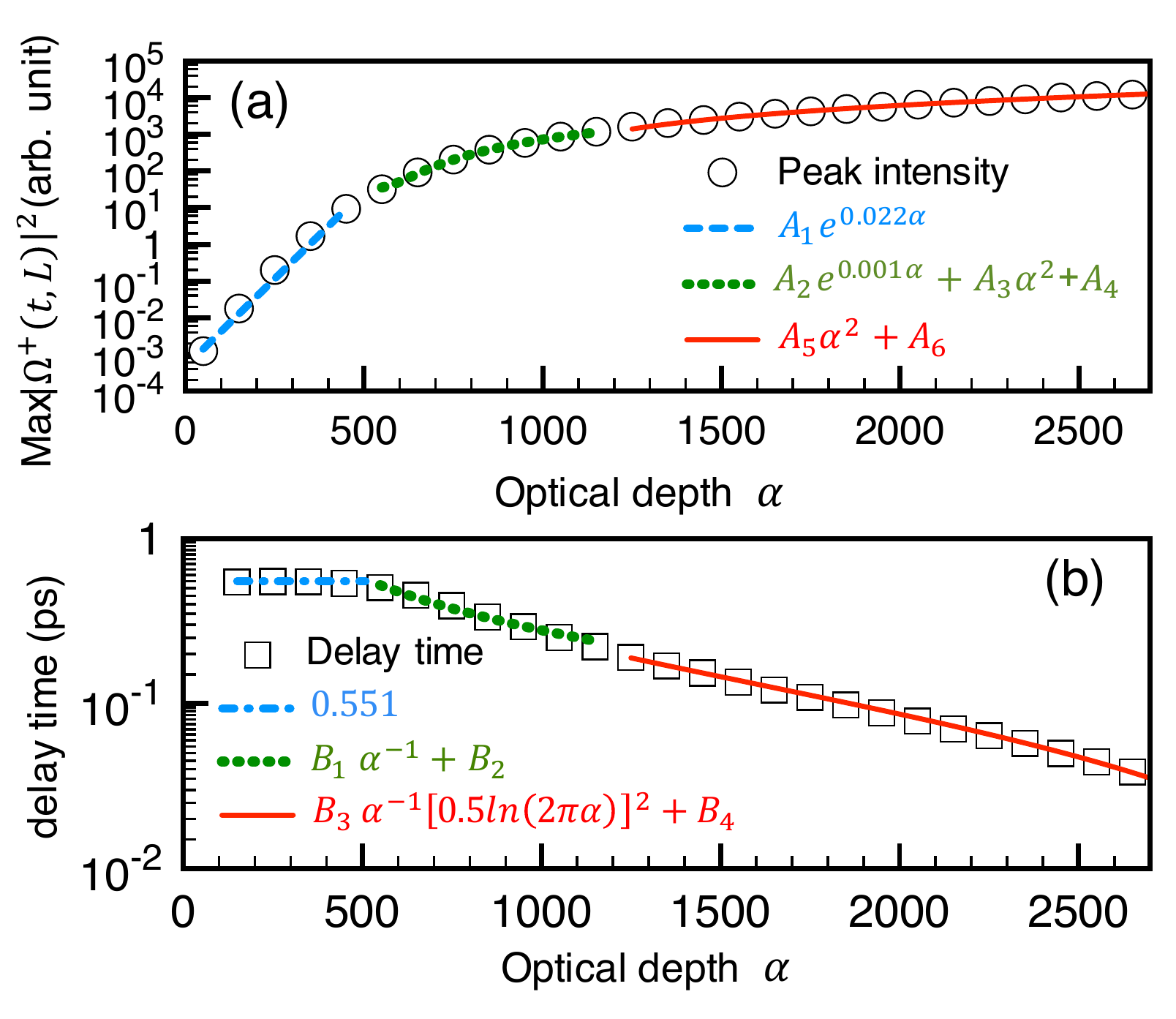}
  \caption{\label{fig8}
(Color online) $\alpha$-dependent (a)  peak intensity and  
(b) delay time of the emitted pulse $\vert\Omega^+(t, L)\vert^2$ for $\tau_2=1$ps. 
Black circles and black squares respectively represent peak intensity and delay time from numerical solutions of $\vert\Omega^+(t, L)\vert^2$.
Each color line illustrates a fitting curve in the correspond $\alpha$ domain.
All parameters are those used in Fig.\ref{fig5}. The fitting coefficients are 
$\left( A_1, A_2, A_3, A_4, A_5, A_6 \right) = $ ( $4.72\times 10^{-4}$, $-5.42\times 10^{3}$, $8.65\times 10^{-3}$, $6.82\times 10^{3}$, $1.96\times 10^{-3}$, $-1.66\times 10^{3}$) and 
$\left( B_1, B_2, B_3, B_4 \right) = $ ( $2.97\times 10^{2}$, $-1.86\times 10^{-2}$, $2.11\times 10$, $-1.5\times 10^{-1}$).
  }
\end{figure}
\begin{figure}[b]
\vspace{-0.4cm}
  \includegraphics[width=0.48\textwidth]{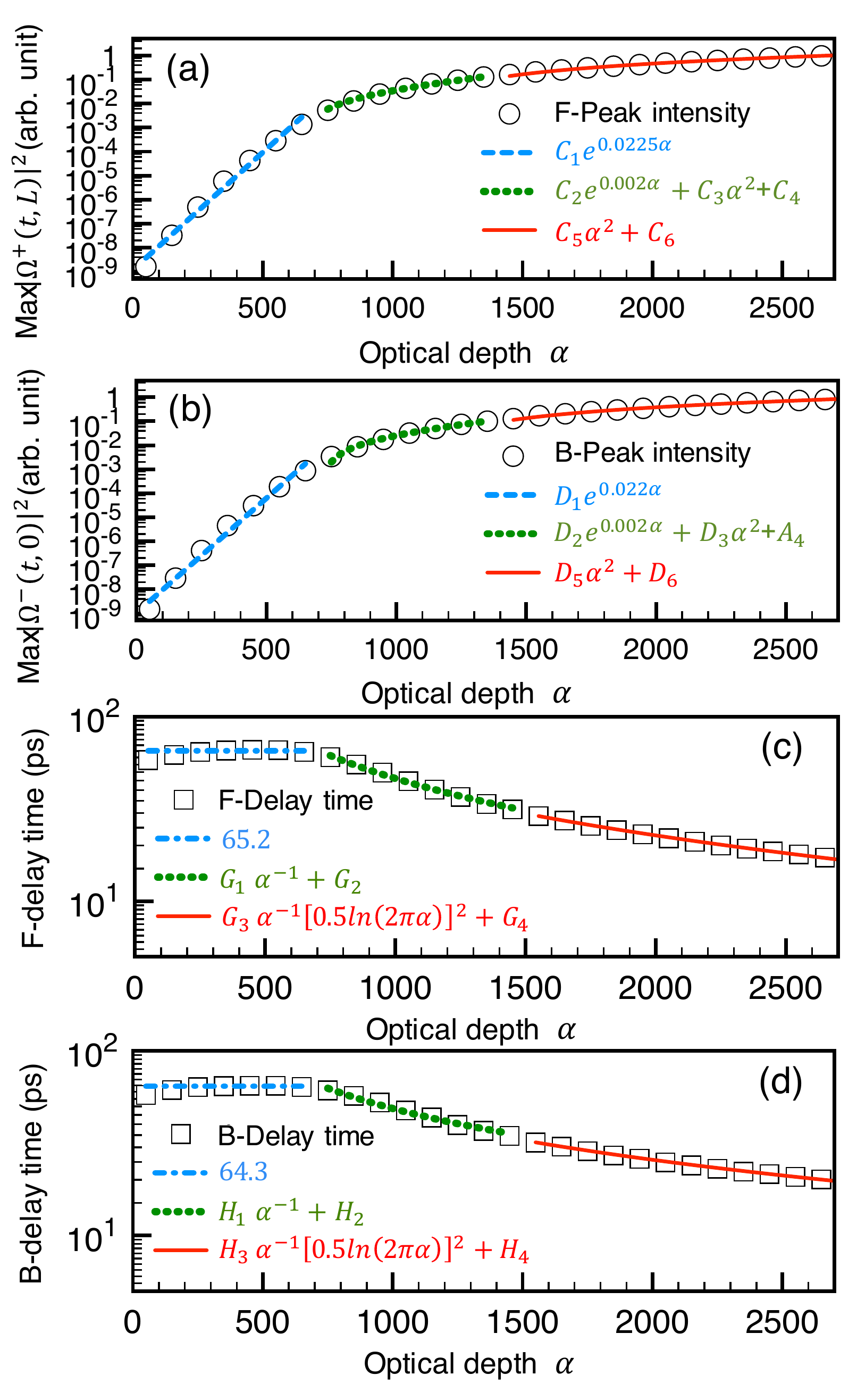}
  \caption{\label{fig9}
(Color online)  $\alpha$-dependent peak intensity of (a) forward $\vert\Omega^+(t, L)\vert^2$ and (b) backward $\vert\Omega^-(t, 0)\vert^2$  for $\tau_2=100$ps.  
$\alpha$-dependent delay time of the emitted pulse (c) $\vert\Omega^+(t, L)\vert^2$ and (d) $\vert\Omega^-(t, 0)\vert^2$. 
Black circles and black squares respectively represent peak intensity and delay time from numerical solutions.
Each color line illustrates a fitting curve in the correspond $\alpha$ domain.
All parameters are those used in Fig.\ref{fig6}. The fitting coefficients are 
$\left( C_1, C_2, C_3, C_4, C_5, C_6 \right) = $ ( $1.21\times 10^{-9}$, $0.021$, $-7.38\times 10^{-8}$, $-0.05$, $1.7\times 10^{-7}$, $-0.22$),
$\left( D_1, D_2, D_3, D_4, D_5, D_6 \right) = $ ( $1.05\times 10^{-9}$, $0.016$, $-5.46\times 10^{-8}$, $-0.04$, $1.41\times 10^{-7}$, $-0.18$),
$\left( G_1, G_2, G_3, G_4 \right) = $ ( $4.58\times 10^{4}$, $3.96\times 10^{-1}$, $2.5\times 10^{3}$, $-5.258$) and
$\left( H_1, H_2, H_3, H_4 \right) = $ ( $4.2\times 10^{4}$, $6.77$, $2.53\times 10^{3}$, $-2.47$).
  }
\end{figure}

In both Fig.~\ref{fig5}(a) and Fig.~\ref{fig6}(a) there are five data sets marked as \textbf{c} ($L$, $n$) = (0.25mm, 2.25$\times 10^{14}$mm$^{-3}$), \textbf{d} (0.5mm, 1$\times 10^{14}$mm$^{-3}$), \textbf{e} (0.5mm, 2.25$\times 10^{14}$mm$^{-3}$), \textbf{f} (0.5mm, 3.5$\times 10^{14}$mm$^{-3}$) and \textbf{g} (0.75mm, 2.25$\times 10^{14}$mm$^{-3}$). 
Average temporal intensity based on Eq.~(\ref{eq17}), average spectral intensity calculated by Eq.~(\ref{eq18}) and photon number histogram of five chosen points are respectively demonstrated by ($\textbf{x}-1$), ($\textbf{x}-2$) and ($\textbf{x}-3$) of Fig.~\ref{fig5} and Fig.~\ref{fig6}, where $\textbf{x}\in \left\lbrace \textbf{c}, \textbf{b}, \textbf{e}, \textbf{f}, \textbf{g} \right\rbrace $.
When scanning parameters through either path \textbf{c}-\textbf{e}-\textbf{g} of lengthening the gain medium with constant density or \textbf{d}-\textbf{e}-\textbf{f} of densifing  the gain medium for a given length,  we observe that the pulse area becomes high,  and the occurrence of optical ringing and spectral splitting becomes significant. 
The typical $\vert\Omega^+\left( t, L\right)\vert^2$ of single realization are  illustrated by the insets of Fig.~\ref{fig5}($\textbf{x}-1$). One can see that the optical ringing effect happens in $\int_{-\infty}^\infty\vert\Omega^+\left( t, L \right) \vert dt\geq\frac{\pi}{2}$ region.
The shortening of  the $\Omega^+(t, L)$ pulse duration is accompanied by the widening of spectral splitting.
The delay time between the peak of $J_p(t, L)$  and the peak of $\Omega^+(t, L)$  also becomes short, namely speed-up emission. 
For a better visualization, we indicate the peak instant of $J_p(t, L)$ by gray dashed lines in Fig.~\ref{fig5}($\textbf{x}-1$), and it is very close to $t=0$ in Fig.~\ref{fig6}($\textbf{x}-1$) of longer time scale.
We demonstrate the time delay histogram of forward emission for $\tau_2=1$ps in the insets of Fig.~\ref{fig5}($\textbf{x}-3$). One can see that the most probable delay time is shifting to small value, and its fluctuation amplitude becomes ten times wide when the probability of $\int_{-\infty}^\infty\vert\Omega^+\left( t, L \right) \vert dt\geq\frac{\pi}{2}$ increases.
Moreover, Fig.~\ref{fig5}($\textbf{x}-3$) and Fig.~\ref{fig6}($\textbf{x}-3$) show that the photon number histogram of the emitted $\Omega^+(t, L)$ spreads from left to the right and gradually peaks at a certain large photon number.
As also revealed by Fig.~\ref{fig5}($\textbf{x}-3$) and Fig.~\ref{fig6}($\textbf{x}-3$), the fluctuation amplitude of photon number is also getting wide when the system goes to high optical depth region.
The $\langle I^{-} \left(t \right) \rangle$ and backward  photon number histogram are  illustrated in Fig.~\ref{fig6}($\textbf{x}-1$)  inset and Fig.~\ref{fig6}($\textbf{x}-3$) inset, respectively. Both share the same tendency with the forward one.
The above features suggest that $\Omega^{\pm}$ experiences a certain transition \cite{Okada1978, Brechignac1981, Boyd1987} around the boundary of  $\int_{-\infty}^\infty\vert\Omega^{+}\left( t, L\right)  \vert dt = \frac{\pi}{2}$ and $\int_{-\infty}^\infty\vert\Omega^{-}\left( t, 0\right)  \vert dt = \frac{\pi}{2}$ which is associated with  Rabi oscillation.
For $\tau_2 = 10$fs, our numerical simulation shows that the population inversion is not produced by $J_p$ when $\tau_p > 30$fs. This confirms the analysis demonstrated in Fig.~\ref{fig4}(a) using the integral approach. Fig.~\ref{fig7} illustrates the results utilizing ($r$, $n_p$, $\tau_p$, $\tau_2$)=(2$\mu$m, $10^{14}$, 24fs, 10fs).
Because the backward emission is negligible in this case, we only depict the probability  of forward emission in Fig.~\ref{fig7}(a).
The three data sets are marked as \textbf{b} ($L$, $n$) = (0.2mm, 6.5$\times 10^{14}$mm$^{-3}$), \textbf{c} (0.3mm, 9.5$\times 10^{14}$mm$^{-3}$) and \textbf{d} (0.44mm, 1.48$\times 10^{15}$mm$^{-3}$), and their corresponding spectra of $\Omega^+(t,L)$ are given in Fig.~\ref{fig7}(b), (c) and (d), respectively.
When moving to the top right along path \textbf{b}-\textbf{c}-\textbf{d}, the spectrum is getting split. This reflects the slightly oscillatory behaviour damped by the high decay rate in the time domain, and the similar transition also happens around  $\int_{-\infty}^\infty\vert\Omega^{+}\left( t, L\right)  \vert dt = \frac{\pi}{2}$ and $\int_{-\infty}^\infty\vert\Omega^{-}\left( t, 0\right)  \vert dt = \frac{\pi}{2}$.

\begin{figure}[b]
\vspace{-0.4cm}
  \includegraphics[width=0.48\textwidth]{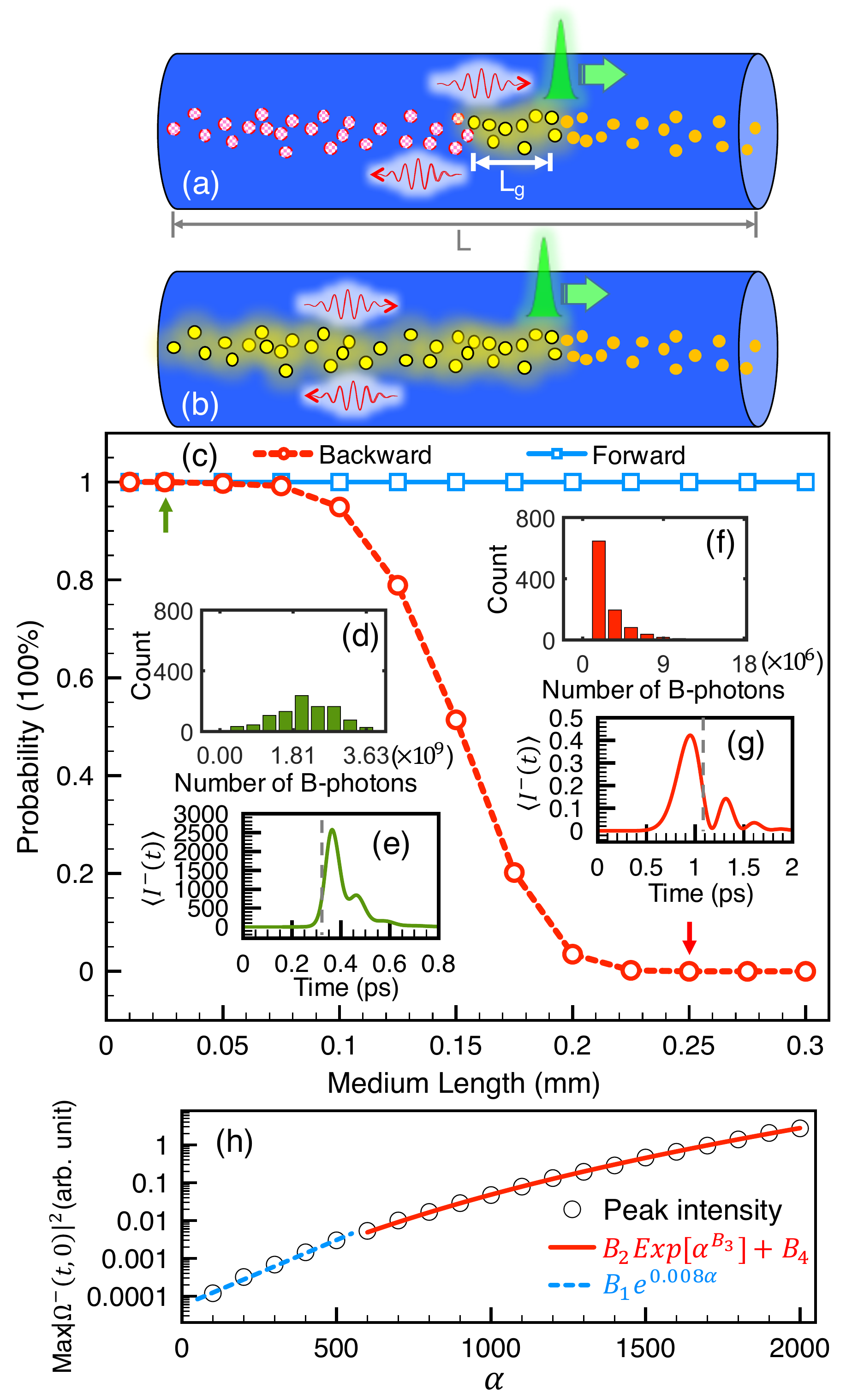}
  \caption{\label{fig2}
(Color online) Forward and backward asymmetry and length induced backward transition. All particles are initially prepared in state  $\vert 0\rangle$ (orange-filled circles).
(a) as $L_g < L$, pumped particles remain in excited state $\vert 2\rangle$ (yellow dots) for only a short distance behind the pumping pulse (green-right-moving Gaussian), and others decay to state $\vert 1\rangle$ (red shaded circles).
(b) case when $L_g > L$.
(c) probability of forward $\int_{-\infty}^\infty\vert\Omega^{+}\left( t, L\right)  \vert dt \geq \frac{\pi}{2}$ (blue solid line) and that of backward $\int_{-\infty}^\infty\vert\Omega^{-}\left( t, 0\right)  \vert dt \geq \frac{\pi}{2}$ (red dashed line) as a function of $L$ among 1000 realizations for $\left( \tau_2, \alpha\right) =$ (1ps, 1500). Other parameters are the same as those used in Fig.~\ref{fig5}. 
(d) backward photon number histogram and (e) $\langle I^{-} \left(t \right) \rangle$ for $L=0.025$mm (indicated by green upward arrow). 
(f) backward photon number histogram and (g) $\langle I^{-} \left(t \right) \rangle$ for $L=0.25$mm (indicated by red downward arrow).
Gray dashed lines indicate instants when $J_{p}$ leaves the medium.
(h) $\alpha-$dependent average backward peak intensity for $L=0.25$mm. Fitting parameters $\left( B_1, B_2, B_3, B_4  \right) = $ ( $5.57\times 10^{-5}$, $6.07\times 10^{-8}$, $3.78\times 10^{-1}$, $4.52\times 10^{-4}$)
  }
\end{figure}
In order to quantify the observed transition,
we collect $\vert\Omega^+(t,L)\vert^2$ and $\vert\Omega^-(t,0)\vert^2$ for each contour  in  Fig.~\ref{fig5}\&\ref{fig6}
and analyse them by finding  the $\alpha$ power law for the averaged peak intensity and for the averaged delay time  between forward/backward emission and the output pumping pulse.  
In Fig.~\ref{fig8}(a) and Fig.~\ref{fig9}(a)(b) the black circles depict the $\alpha$-dependent peak intensity. 
In Fig.~\ref{fig8}(b) and Fig.~\ref{fig9}(c)(d)
the black squares demonstrate the $\alpha$-dependent delay time.  The blue dashed lines are fitting curves for $\alpha\leq 500$, green dotted line for $500 < \alpha < 1200$ and red solid lines for $\alpha > 1200$. 
In Fig.\ref{fig8} and Fig.\ref{fig9} the domain $500 < \alpha < 1200$ and $700 < \alpha < 1500$ are obviously the watershed of power law, respectively. 
To the left of the watershed the peak intensity exhibits exponential growth with a constant delay time of 0.551 ps for $\tau_2 = 1$ps and that of about 65ps for $\tau_2 = 100$ps, which typically results from ASE.
The upper bound of the ASE gain exponent of 0.055 and that of 0.076 are given by Eq.~(\ref{eq31}) for $\tau_2 = 1$ps and $\tau_2 = 100$ps, respectively.
Both are in the same order of magnitude with  the fitting value of 0.022 (blue dashed line in Fig.~\ref{fig8}(a) and Fig.~\ref{fig9}(a)(b)).
However, to the right of the watershed the peak intensity is proportional to $\alpha^{2}$, and the delay time behaves as the typical form of $\tau_D\propto\alpha^{-1}\left[ \frac{1}{2}\ln\left( 2\pi\alpha\right) \right] ^2$ \cite{Rehler1971, Polder1979, Gross1982}.
The former corresponds to the collective emission and the latter reflects the needed time for building up coherence from noise \cite{Polder1979, Gross1982},  which are both signatures of superfluorescence. 
It is worth mentioning that the typical form of superfluorescence delay time deviates from the numerical values when $\alpha < 1200$ and $\alpha < 1500$ for $\tau_2 = 1$ps and $\tau_2 = 100$ps, respectively.
In the watershed domain, the green dotted line in Fig.~\ref{fig8}(a) and Fig.~\ref{fig9}(a)(b) indicate that the peak intensity behaves as a superposition of exponential growth and $\alpha^{2}$. This tendency reveals that a transition does happen from one emission mechanism to another \cite{Okada1978, Boyd1987}.
The present results also suggest that the optical depth $\alpha$ plays the key role for a  transition from ASE to SF. This transition happens at the boundary  where Rabi oscillation also starts to occur. 
The validity of one dimensional simulation is associated with the Fresnel number condition $\mathbf{F} = \frac{\pi r^2}{L \lambda}\approx 1$.
Otherwise one has to deal with full three-dimensional wave propagation in simulations for the transverse effect of diffraction \cite{Gross1982}, which  consumes a lot of computational time and power especially for the ensemble average. To avoid such heavy computation, we have compared averaged results due to different values of $n$ and $L$ for $n L=$ constant, i.e., $\alpha =$ constant, in our model and do not observe any significant difference. Therefore, one could take the advantage of the $\alpha$-dependent  features and effectively use a system which  fulfils the Fresnel  condition with the same $\alpha$.

Given that the boundary between ASE and SF may occur when $\int_{-\infty}^\infty\Omega\left( t\right)  dt=\frac{\pi}{2}$, we here estimate the photon number of an emitted $\frac{\pi}{2}$-pulse. For simplicity, we use a Gaussian pulse such that $\int_{-\infty}^\infty\Omega Exp\left[ -\left( \frac{t}{\kappa\tau_2} \right)^2 \right] dt\approx\Omega\sqrt{\pi}\kappa\tau_2 = \frac{\pi}{2}$  where $\kappa\tau_2$ is the duration of the emitted pulse, and $\kappa=\ln 2$ given by the duration of positive population inversion as illustrated in Fig.~\ref{fig3}. One can therefore obtain $\Omega \approx \frac{\sqrt{\pi}}{2\tau_2\ln 2}$.
In view of the fact that the integral of laser intensity equals the energy of $n_e$ photons,
 $\frac{1}{2}c\varepsilon_0 \pi r^2\int_{-\infty}^\infty\frac{\hbar^2}{d^2}\vert\Omega\left( t\right) \vert^2 dt = n_e \hbar\omega$, which results in
$n_e \approx \frac{\hbar c \varepsilon_0 \pi^{5/2} r^2}{8\sqrt{2} \omega  d^2\ln 2}\Gamma$. 
By substituting $c = \omega\lambda/\left( 2\pi \right) $ and the spontaneous decay rate
$
\Gamma = \frac{ d^2 \omega^3}{3\pi\varepsilon_0 \hbar c^3},
$
we obtain
\begin{equation}
n_e \approx \frac{\pi^\frac{7}{2} r^2}{6\sqrt{2}\lambda^2\ln 2},
\end{equation}
which
suggests that $n_e$ of a $\frac{\pi}{2}$-pulse mainly depends on  the ratio of $r$ to $\lambda$. 
In the present cases, $n_e = 1.75\times 10^7$  is very close to our numerical result of $3.24\times 10^7$. 

\begin{figure}[b]
\vspace{-0.4cm}
  \includegraphics[width=0.45\textwidth]{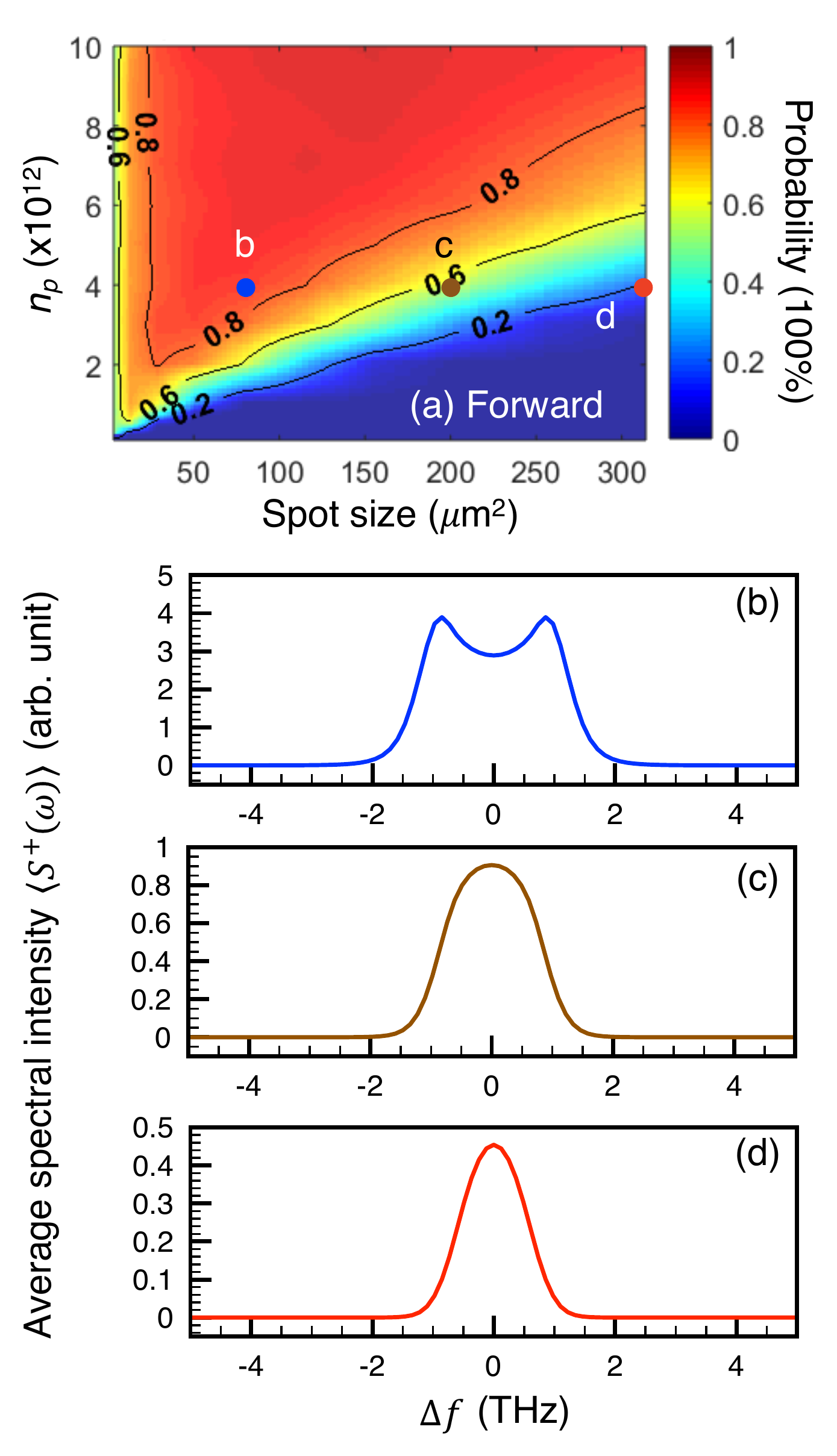}
  \caption{\label{fig10}
(Color online) (a) probability of superfluorescence occurs as a function of ($\pi r^2$, $n_p$) in  forward direction among 1000 realizations for $\left(  L, n, \tau_2\right) = \left(  0.16 \mathrm{mm}, 5\times 10^{14} \mathrm{mm^{-3}}, 1 \mathrm{ps}\right)$ with Gaussian random noise. 
Three data sets, marked as \textbf{b} (78.54$\mu$m$^2$, 4$\times 10^{12}$), \textbf{c} (201.06$\mu$m$^2$, 4$\times 10^{12}$), \textbf{d} (314.16$\mu$m$^2$, 4$\times 10^{12}$) in (a), are chosen for showing the transition from amplified spontaneous emission to superfluorescence. The corresponding average spectra are respectively illustrated in (b), (c) and (d).
  }
\end{figure}
\section{Length-Induced Backward Transition}\label{sec:fba}
We turn to investigate the  forward-backward asymmetry demonstrated in Fig.~\ref{fig5}(a)(b) and its relation with  $L_g$ and $L$.  
Fig.~\ref{fig2}(a) shows that regions sandwich the small gain  area (filled with yellow  shaded circles)  causes an asymmetric environment for light propagating forwards and backwards. For $L_g < L$ there is always a gain medium ($\rho_{22}>\rho_{11}$) ahead for light propagating forward, but  an absorption medium ($\rho_{22}<\rho_{11}$)  for the backward radiation. This is one of reasons why backward emission is often absent in a swept pumping system. In contrast, Fig.~\ref{fig2}(b) reveals that choosing a system with $L_g > L$ will lead to symmetric behaviour  for both directions. As a result, the increase of $\Omega^-$ gain and events of $\int_{-\infty}^\infty\vert\Omega^-\left( t, 0\right)  \vert dt\geq\frac{\pi}{2}$ are observed for $\tau_2=100$ps in  Fig.~\ref{fig6} (b).

Given $L_g=0.15$mm for $\tau_2=1$ps as demonstrated in Fig.~\ref{fig3}(a), the backward superfluorescence is  anticipated to show up as $L<0.15$mm but not when $L>0.15$mm.
In what follows we use constant $\alpha=1500$ and same parameters used in  Fig.~\ref{fig5} to demonstrate length induced backward ASE-SF transition in Fig.~\ref{fig2}(c).
The blue solid line and red dashed line respectively represent the probability  of $\int_{-\infty}^\infty\vert\Omega^+\left( t, L \right) \vert dt\geq\frac{\pi}{2}$  and that of $\int_{-\infty}^\infty\vert\Omega^-\left( t, 0 \right) \vert dt\geq\frac{\pi}{2}$  among 1000 realizations of simulation for each length. 
Fig.~\ref{fig2}(d) and (e) show backward photon number histogram and $\langle I^{-} \left(t \right) \rangle$, respectively for $L=0.025$mm, and Fig.~\ref{fig2}(f)(g) for $L=0.25$mm.  $J_{p}$ leaves the medium at the instants pointed out by gray dashed lines.
As one would expect from Fig.~\ref{fig5}(a) that forward probability remains 100\% for the whole length range. However, the backward emission reveals five interesting features when shortening $L$ across the critical value of $L_g=0.15$mm:
(1) probability noticeably raises to 100\%;
(2) $\langle I^{-} \left(t \right) \rangle$ catches up $J_p$;
(3) backward light pulse exits the medium earlier than $J_p$ does when $L\gtrsim L_g$ as depicted by Fig.~\ref{fig2}(g);
(4) photon number histogram demonstrates similar ASE-SF transition as depicted in Fig.~\ref{fig5} and \ref{fig6};
(5) $\langle I^{-} \left(t \right) \rangle$ manifests optical ringing as illustrated by Fig.\ref{fig2}(g) (also in single cases).
Above features support our picture of the length effect and are consistent with  Fig.~\ref{fig3}(a) and Eq.~(\ref{eq25}) which is the consequence of $J_p\left( t, 0\right) $ and $\tau_2$. 
The tiny pulse peaks at $t=1.3$ps in Fig.\ref{fig2}(g) suggests that the backward emission from region deeper than $L_g$ is possible. In view that the backward pulse area for $L \gtrsim L_g$ in Fig.~\ref{fig2}(c) is mostly smaller than $\pi/2$, one would expect it is in the low gain and linear region.
However, the optical ringing is never observed in the ASE region of Fig.\ref{fig6}, and so
feature (5) raises the following question for backward emission.  Why can small pulse area and optical ringing  coexist in the range of  $L_g\sim L$ but cannot in $L_g > L$? 
Since the backward pulse duration in Fig.~\ref{fig2}(g)  is shorter than  $\tau_2$ limited by $L_g$, a broadband small-area pulse envelope should also oscillate  when propagating through a resonant medium of high optical depth \cite{Crisp1970}.
This will not happen to forward ASE because the gain medium co-moves  along with it, making forward ASE duration always comparable to $\tau_2$, i.e., narrowband, as demonstrated in Fig.~\ref{fig5} and \ref{fig6}.
Nevertheless, the non-oscillating ASE in Fig.~\ref{fig6} suggests there is other mechanism than Ref.\cite{Crisp1970}, namely, attenuating SF.
Because $\alpha =1500$ is very high, the backward emission can quickly grow  into SF in the gain region and then enters the dissipative but still penetrable area such that small pulse area and optical ringing  can coexist in the range of  $L_g\sim L$.
In Fig.~\ref{fig2}(h)  we depict average backward peak intensity as function of $\alpha$ by varying $n$ but fixing $L=0.25$mm. Each point is also an average over 1000 realizations of simulation.
The fitting  shows that $\Omega^-\left( t, 0\right) $ do experience a transition at around $\alpha =500$ and manifests   nonlinear evolution lower than $\alpha^2$ dependence.
In contrast, the forward light behaves identically as that in Fig.~\ref{fig8}.
Similar study for $L=0.025$mm also shows identical ASE-SF transition for both forward and backward emissions.
Our backward study suggests that (I) a triggered small-area ASE by an external seeding pulse shorter than $\tau_2$ may lead to also an oscillating ASE in both forward and backward  direction due to the mechanism of Ref.\cite{Crisp1970}; and (II) the transverse pumping may ease the forward-backward asymmetry.

\section{Transition induced by XFEL $J_p$ Laser parameters}\label{sec:Jptransition}
It is a natural question to ask whether  one could manipulate  $\Omega^{+}$ by changing $J_p$ laser  \cite{Skribanowitz1973, MacGillivray1976, Cui2013, Cui2017} based on XFEL parameter \cite{EuropeanXFEL}. In Fig.~\ref{fig10}(a) we demonstrate the probability of the occurrence of $\int_{-\infty}^\infty\vert\Omega^{+}\left( t, L \right) \vert dt\geq\frac{\pi}{2}$ as a function of $(\pi r^2, n_p)$. In our simulation, the pumping photon flux is affected by both $r$ and $n_p$ as denoted by Eq.~(\ref{eq13}), and the Gaussian white noise also depends on $r$ as indicated by Eq.~(\ref{eq16}).
Given the short lifetime $\tau_2 = 1$ps,  gain growth only happens to the forward emissions. 
We use $\left(  L, n, \tau_2\right) = \left(  0.16 \mathrm{mm}, 5\times 10^{14} \mathrm{mm^{-3}}, 1 \mathrm{ps}\right)$ to numerically solve Eq.(\ref{eq1}-\ref{eq16}) for 1000 realizations of simulation at each combination of $J_p$ laser spot size and photon number. The three data sets are marked as \textbf{b} (78.54$\mu$m$^2$, 4$\times 10^{12}$), \textbf{c} (201.06$\mu$m$^2$, 4$\times 10^{12}$) and \textbf{d} (314.16$\mu$m$^2$, 4$\times 10^{12}$) in Fig.~\ref{fig10}(a).
Their averaged spectra based in Eq.~(\ref{eq18}) is respectively demonstrated in Fig.~\ref{fig10}(b), (c) and (d).
When scanning parameters through the \textbf{b}-\textbf{c}-\textbf{d} path of shrinking $J_p$ spot size and fixing pumping photon number, the occurrence of spectral splitting  also becomes obvious. As a consequence, one can manipulate the properties of emitted $\Omega^{\pm}$  not only by changing the parameters of the gain medium  but also by altering that of  pumping laser $J_p$.

\section{Summary}\label{sec:summary}
We have demonstrated the transition between ASE and SF in a three-level-$\Lambda$ type system induced by the change of optical depth of a medium and by the alternation of pumping XFEL parameters, namely, focus spot size and pulse energy. 
A consistent picture of the transition from one region to another suggested by the Maxwell-Bloch equation is summarized in what follows. A pencil-shape gain medium is longitudinally and incoherently pumped by a short XFEL pulse, and then the inverted medium experiences spontaneous decay. Although the spontaneously emitted photons go to all directions, a small number of forward emitted photons follow the XFEL pulse  and enter the solid angle $\varphi$, as demonstrated in Fig.~\ref{fig1}, within which photons may subsequently interact with other excited particles.
The backward emitted photons may  also interact with the inverted particles or may be reabsorbed by the pre-decayed particles behind the gain region depending on $L_g \geq L$ or $L_g < L$, respectively (see Fig.~\ref{fig2}).
%
%
Due to the geometry, the forward emitted light and the backward one, in the above former case,  are both amplified along the long axis of the gain medium. As the pulse area approaches $\pi/2$, the ASE-SF transition starts to occur and results in, e.g., optical ringing effect, spectral splitting  and the change in statistics behaviours as  demonstrated in Fig.~\ref{fig5}, ~\ref{fig6}, \ref{fig8} and \ref{fig9}.
We have investigated the pumping procedure in detail using XFEL parameters and identified $L_g$ and $L$ as two key parameters making forward-backward asymmetry. Moreover, in the region of $L_g\gtrsim L$,we have also studied the length-induced backward transition for the first time.
The present results demonstrate a controllable single-pass light source whose properties can be manipulated by parameters of a medium or those of a pumping XFEL.
\section{Acknowledgement}
We thank Jason Payne for carefully reading our manuscript.
Y.-H. K. and W.-T. L. are supported by the Ministry of Science and Technology, Taiwan (Grant No. MOST 107-2112-M-008-007-MY3 and Grant No. MOST 107-2745-M-007-001-). 
W.-T. L. is also supported by the National Center for Theoretical Sciences, Taiwan.


\bibliographystyle{apsrev}
\bibliography{20191023SF}

\end{document}